\documentclass[12pt]{iopart}

\pdfoutput=1

\usepackage{color}
\usepackage{cite}
\usepackage{url,psfrag,graphicx}
\usepackage{dcolumn}
\usepackage{bm}
\usepackage{hyperref}
\usepackage{float,epsfig,color}
\usepackage{times}
\usepackage{array}
\usepackage{braket}
\usepackage{diagbox}
\usepackage{lscape}
\usepackage{multirow}
\usepackage{stackrel}
\usepackage{verbatim}
\usepackage{xcolor}
\usepackage{soul}
\usepackage{hyperref}
\usepackage{iopams}
\usepackage{setstack}

\newcommand{\eqref}[1]{(\ref{#1})}

\DeclareMathAlphabet      {\mathitbf}{OML}{cmm}{b}{it}

\begin{document}

\title[Numerical integration of the KPZ and related equations on networks]{Numerical integration of the KPZ and related equations on networks: the case of the Cayley tree\footnote{Published in \textit{J.\ Stat.\ Mech.: Theor. Exp.} \textbf{2025} 083203.}}

\author{J. M. Marcos$^{1,2}$, J. J. Mel\'endez$^{1,2}$, R. Cuerno$^3$, J. J. Ruiz-Lorenzo$^{1,2}$}

\address{$^1$ Departamento de  F\'{\i}sica, Universidad de Extremadura, 06006 Badajoz, Spain}
\address{$^2$ Instituto de Computaci\'on Cient\'{\i}fica Avanzada de Extremadura (ICCAEx), Universidad de Extremadura, 06006 Badajoz, Spain}
\address{$^3$ Universidad Carlos III de Madrid, Departamento de Matemáticas and Grupo Interdisciplinar de Sistemas Complejos (GISC), Avenida de la Universidad 30, 28911 Leganés, Spain}

\date{\today}

\begin{abstract}

The numerical integration of stochastic growth equations on non-Euclidean networks presents unique challenges due to the nonlinearities that occur in many relevant models and to the structural constraints of the networks. In this work, we integrate the Kardar–Parisi–Zhang (KPZ), Edwards-Wilkinson, and tensionless KPZ equations on Cayley trees using different numerical schemes and compare their behavior with previous results obtained for discrete growth models. By assessing the stability and accuracy of these methods, we explore how network topology influences interface growth and how boundary effects shape the observed scaling properties. Our results show good agreement with previous studies on discrete models, reinforcing key scaling behaviors while highlighting some differences. These findings contribute to a better understanding of surface growth on networked substrates and provide a computational framework for studying nonlinear stochastic processes beyond Euclidean lattices.

\end{abstract}

\section{Introduction}\label{sec:intro}

The Kardar-Parisi-Zhang (KPZ) equation \cite{Kardar1986},
\begin{equation}
    \frac{\partial h}{\partial t}\left(\boldsymbol{x},t\right)=\nu \nabla^2 h\left(\boldsymbol{x},t\right)+\frac{\lambda}{2}\left(\nabla h\right)^2\left(\boldsymbol{x},t\right)+\eta\left(\boldsymbol{x},t\right)\, ,
    \label{eq:kpz}
\end{equation}
characterizes the macroscopic scale-invariant behavior of many non-equilibrium growth processes \cite{Barabasi1995,Krug1997,Takeuchi2018}, where $h\left(\boldsymbol{x},t\right)$ represents the height value of an evolving interface above position $\boldsymbol{x}\in \mathbb{R}^d$ on a reference substrate, at time $t$. The first term on the right-hand side accounts for the interface relaxation driven by surface tension $\nu$, the second term implements growth at a constant rate along the local normal direction to the surface, and the last term corresponds to an uncorrelated Gaussian white noise, with zero-average, $\langle \eta\left(\boldsymbol{x},t\right)\rangle=0$, and correlations $\langle \eta\left(\boldsymbol{x},t\right)\eta\left(\boldsymbol{x'},t'\right)\rangle=2D\delta^d\left(\boldsymbol{x}-\boldsymbol{x'}\right)\delta\left(t-t'\right)$.

Although surface growth phenomena remain among the most relevant applications of the KPZ equation, this continuum model is recently becoming relevant for many other instances of nonequilibrium physics, from active \cite{Caballero2020} to quantum \cite{Sieberer2025} matter. Furthermore, this equation is directly related to other interesting continuum models associated with universality classes that differ from that of KPZ; namely, when $\lambda = 0$, the equation reduces to the linear Edwards-Wilkinson (EW) equation \cite{Edwards1982,Barabasi1995}, which coincides with the Gaussian approximation to the stochastic Ginzburg-Landau equation at criticality \cite{Kardar2007}. Likewise, setting $\nu = 0$ yields the tensionless KPZ (TKPZ) equation \cite{Cartes2022,RodriguezFernandez2022}, potentially relevant for, e.g., percolation-related growth processes \cite{RodriguezFernandez2022} or for reaction-diffusion systems displaying phase turbulence \cite{Vercesi2024}. 

The space-time scale invariance properties of these equations are largely encoded into the so-called Family-Vicsek (FV) dynamic scaling ansatz for the global roughness or surface width $w(L,t)$ (namely, the root-mean-square deviation of height fluctuations around their space average)
\cite{Barabasi1995,Krug1997},
\begin{equation}
	\label{eq:w}
	w(L,t)=t^{\beta}f\left( t/L^z \right),
\end{equation}
with $f(\cdot)$ a suitable scaling function so that the roughness increases as $w\sim t^{\beta}$ for short times such that $t \ll L^z$, while it saturates to a time-independent value  $w_{\mathrm{sat}}\sim L^{\alpha}$ for long times such that $t \gg L^z$. Here, $\alpha$ is the roughness exponent, which is related with the fractal dimension of the surface or interface \cite{Barabasi1995,MozoLuis2022} and characterizes its height fluctuations or roughness at (large) scales comparable with the system size $L$. Further, $\beta$ in Eq.~\eqref{eq:w} denotes the growth exponent and $z$ is the so-called dynamic exponent, which quantifies the power-law increase of the lateral correlation length $\xi(t)\sim t^{1/z}$ along the front. In the FV ansatz the exponents satisfy the relation $\alpha=\beta z$ \cite{Barabasi1995,Krug1997}.

In the case of the Edwards-Wilkinson equation the exponents can be calculated analytically for all substrate dimensions $d$ \cite{Barabasi1995,Kardar2007}:
\begin{equation}
	\label{eq:exponentes_ew}
	\alpha=\frac{2-d}{2},\hspace{2mm}\beta=\frac{2-d}{4},\hspace{2mm}z=2.
\end{equation}
For $d\leq 2$, the width satisfies the FV scaling ansatz. Right at $d=2$, the width scales logarithmically with time at short times before saturation, while its (squared) saturation value scales logarithmically with the system size. For $d>2$ the roughness exponent becomes negative and the surface is flat \cite{Barabasi1995}, in the sense that the roughness no longer scales with the system size $L$. Thus, the upper critical dimension $d_u$ of the EW universality class is $d_u=2$, indeed as for the Gaussian approximation to the Ginzburg-Landau model \cite{Kardar2007}.

For the KPZ equation the exponents can be calculated analytically only in the one-dimensional case, these being \cite{Barabasi1995,Krug1997,Takeuchi2018}
\begin{equation}
	\label{eq:exponentes_kpz_1d}
	\alpha=1/2,\hspace{2mm}\beta=1/3,\hspace{2mm}z=3/2,
\end{equation}
while the so-called Galilean scaling relation $\alpha + z = 2$ is expected to hold at the KPZ fixed point (in the renormalization group (RG) sense \cite{Kardar2007}) for all $d\leq d_u$, so that only one of the exponents remains independent. In higher dimensions $d>1$, the exponents can be computed by integrating the equation numerically 
or by e.g.\ simulating discrete growth models in the same universality class, see e.g.\ Refs.\ \cite{Barabasi1995,Takeuchi2018,Oliveira2022} and other therein. 
In general, the nonlinear behavior of the KPZ equation is controlled by the coupling $g=\lambda^2 D/\nu^3$ \cite{Amar1990,Hentschel1991,Barabasi1995}, becoming more prevalent e.g.\ when $\lambda$ and/or $D$ increase, of if $\nu$ decreases. The value of $g$ can actually be an indicator of the numerical stability of the equation, which decreases for a stronger nonlinearity \cite{Gallego2011}. Furthermore, the KPZ equation presents a non-equilibrium roughening transition transition in terms of the value of $g$, between a smooth phase and a rough phase \cite{Tang1990,Barabasi1995,Krug1997,Tauber2014}. For $d < 2$ the KPZ equation is always in the rough phase, displaying the traits of the KPZ universality class. However, for $d>2$ the rough phase can be accessed only if the coupling parameter is larger than a certain critical value, $g > g_c$. When $g<g_c$ the non-linear term is irrelevant, and the scaling behavior is that of the EW universality class in the corresponding dimension. This regime is known as the weak coupling regime. When $g>g_c$, the nonlinear term is relevant and KPZ scaling ensues. The critical value $g_c$ grows with the dimension of the system, with $d=2$ being the lower critical dimension for the transition. 

In the case of the tensionless KPZ equation, significantly less is known about its properties. This equation is marginally unstable to periodic perturbations of a flat solution \cite{Cuerno1995}, making its numerical integration particularly difficult \cite{Tabei2004,Bahraminasab2004}. Only recently has the numerical integration of the equation become feasible \cite{Cartes2022,RodriguezFernandez2022}. The TKPZ equation defines a distinct universality class, characterized by critical exponents that differ from those of the standard KPZ equation \cite{RodriguezFernandez2022}. Besides, the dynamic scaling ansatz that ensues differs from FV, being termed intrinsically anomalous scaling \cite{Lopez1997, Ramasco2000}. In $ d = 1 $ \cite{RodriguezFernandez2022}, the roughness and dynamic exponents computed numerically take the values $ \alpha = z = 1 $, consistent with the KPZ Galilean scaling relation, while the growth exponent is $ \beta = 1 $ \cite{RodriguezFernandez2022}. In presence of intrinsic anomalous scaling, the roughness exponent that characterizes height fluctuations at local scales, $\alpha_{\mathrm{loc}}$, differs from that one ($\alpha$) characterizing fluctuations at large scales comparable to the system size \cite{Lopez1997, Ramasco2000}. Indeed, for the 1D TKPZ equation the computed local roughness exponent is $ \alpha_{\mathrm{loc}} = 1/2 \neq \alpha$ \cite{RodriguezFernandez2022}. A tensionless or inviscid fixed point with $z=1$ has been identified within the RG structure of the KPZ equation \cite{Fontaine2023}, and is believed to occur for any $d>1$ with $d$-independent exponents \cite{Gosteva2024}.

In general, as in equilibrium critical systems \cite{Kardar2007,Tauber2014}, scaling exponents are expected to depend on $d$ for the KPZ class as well, as long as $d< d_u$; indeed, there is a body of numerical and theoretical work aimed at obtaining the KPZ critical exponents as functions of the system dimension, see e.g.\ Refs.\ \cite{Barabasi1995,Alves2014,Oliveira2022} and other therein. In parallel with the results, Eq.~\eqref{eq:exponentes_ew}, for the Gaussian approximation of KPZ (or EW class), for the KPZ universality class the critical exponents are expected to show logarithmic corrections at $d=d_u$, while for $d>d_u$ diffusive behavior ($z=2$) and lack of scaling with system size (i.e., $\alpha=\beta=0$) are expected. Indeed, the reader can find a summary of the values of the growth exponent obtained from different simulations up to $d=15$ in the recent Ref.\ \cite{Oliveira2022}, together with a formula for $\beta(d)$ which, by continuously decreasing towards 0, shows good overall agreement with simulation results, although recent accurate results~\cite{HalpinHealy2025} show some deviations from this prediction. However, to date the occurrence and precise value of $d_u$ for KPZ remain to be unambiguously determined. Analytically, there are different predictions for $d_u$, some predicting that $d_u<4$ \cite{Halpin-Healy1990,Bouchaud1993,Doherty1994,Colaiori2001}, others predicting that $d_u>4$ \cite{Kloss2014,Kloss2014-2}, and others that $d_u=\infty$ \cite{Castellano1998,Castellano1998-2}. From simulations of different models in the KPZ class, there is strong evidence that, if $d_u$ is finite, it is not small, as it has never been encountered in simulations up to $d_u>15$ \cite{Kim2014,Alves2014,Alves2016,Oliveira2022}.

Prior to our work, Saberi \cite{Saberi_2013} simulated some models of the KPZ universality class on the Bethe lattice in order to investigate the KPZ upper critical dimension. The Bethe lattice has been used as an approximation of infinite-dimensional systems in some cases. Due to its distinctive topological structure, several statistical models involving interactions defined on the Bethe lattice are exactly solvable \cite{Baxter2016}. For instance, the Ising model is one such case, featuring the same critical exponents as in the mean-field approximation \cite{Kurata1953}; a similar behavior is found for isotropic percolation \cite{Christensen2005}. In Ref.\ \cite{Saberi_2013}, Saberi showed that the roughness of some discrete growth models follows logarithmic scaling and thus he claimed that the KPZ nonlinearity was still relevant in infinite dimensions, denying the existence of a finite upper critical dimension for the KPZ universality class. More recently, Oliveira \cite{Oliveira2021} revisited the work of Saberi concluding that the standard deviation of a non-flat surface had been measured in \cite{Saberi_2013}, incorrectly reporting this as the surface roughness. Oliveira argued that when the surfaces are non-flat the height fluctuations have to be measured for a single or a few surface points \cite{Alves2013,Saberi2019,Oliveira2013} due to the loss of the spatial translation symmetry.

A natural alternative to these works on discrete models is to numerically integrate the KPZ equation itself in high-dimensional systems. Note at this that many numerical simulations of the KPZ equation in regular lattices use the simplest explicit discretization of the squared gradient \cite{Amar1990,Moser1991}, namely (for $d=1$, for simplicity),
\begin{equation}
	\label{eq:grad_discr}
	(\nabla h_i)^2\approx (h_{i+1}-h_{i-1})^2,
\end{equation}
where the notation $i\pm1$ refers to the right-left neighbors of a given site $i$. The discrete nonlinear term, Eq.\ \eqref{eq:grad_discr}, is highly unstable and, for large values of the nonlinear parameter $\lambda$, the numerical integration of Eq.\ \eqref{eq:kpz} diverges when some large fluctuation introduced by the noise grows faster than the Laplacian can relax away \cite{Dasgupta1996,Dasgupta1997}. This instability is an intrinsic property of the discretized KPZ equation, arising due to isolated pillars that quickly grow in time; the rapid growth of these pillars leads to numerical blow up when the coupling constant exceeds a critical value \cite{Dasgupta1997}. 
To overcome these difficulties, some improvements have been proposed, such as improved real-space discretizations of the nonlinear term \cite{Lam1998} or pseudospectral schemes \cite{Giada2002,Gallego2011,Toral2014}. While the pseudospectral approximation requires periodic boundary conditions (BC) and thus it is not possible to adapt it to a network like the Cayley tree, the discretization proposed by Lam and Shin \cite{Lam1998} can be redefined in any network. This is the approach that will be adopted in our present work.

In addition, another way to control the intrinsic instability of the discretized equation, that has been used with great success for the KPZ and for other kinetic roughening universality classes \cite{Dasgupta1996,Dasgupta1997,Miranda2008,Ales2019,Song2021}, is to replace the term $(\nabla h_i)^2$ in the discretized equation by $f((\nabla h_i)^2)$, where $f(x)=(1-e^{-cx})/c$, with $c>0$ being an adjustable parameter. This replacement corresponds to the introduction of an infinite series of higher powers of $(\nabla h_i)^2$, whose coefficients depend on the value of $c$. When $c$ is larger than a critical value, the instability is completely suppressed from the discretized equation, enabling reliable estimates of scaling exponents \cite{Dasgupta1996,Dasgupta1997,Miranda2008,Ales2019,Song2021}, while the growth equation may be numerically unstable for $c=0$ when such terms are not included. This approach will also be used in this paper.

In this work we numerically integrate the KPZ equation on finite approximations of the Bethe lattice, i.e., on Cayley trees, and compare the results with those obtained by Saberi \cite{Saberi_2013} and Oliveira \cite{Oliveira2021} for discrete models in the KPZ class. We aim to push further the exploration of the KPZ upper critical dimension by simulating one more member of the universality class, namely, the continuum equation itself. In our study we will vary the value of the $\lambda$ parameter of the nonlinear term in the equation, from $\lambda =0$ in the EW Gaussian approximation, up to very large effective values (namely, $\nu=0$) as for the TKPZ equation. This is at variance with the discrete models previously studied, for which the effective value of $\lambda$ in their putative continuum descriptions cannot be independently tuned, and is hoped to provide indications on the effect of the nonlinearity in the high-dimensional substrates we will employ.

This paper is organized as follows. Section \ref{observables} contains a description of the numerical integration methods used in this work and the definition of the various observables measured. Our numerical results are then reported and discussed in Sec.\ \ref{sec:results}. A summary of our results (including their compact collection into a table for the reader's convenience) closes that section, followed by our conclusions in Sec.\ \ref{sec:concl}. Finally, a number of additional numerical results are collected into four appendices at the end.

\section{Definitions and simulation details} \label{observables}

\begin{figure}[t]
\centering
\includegraphics[width=0.495\textwidth]{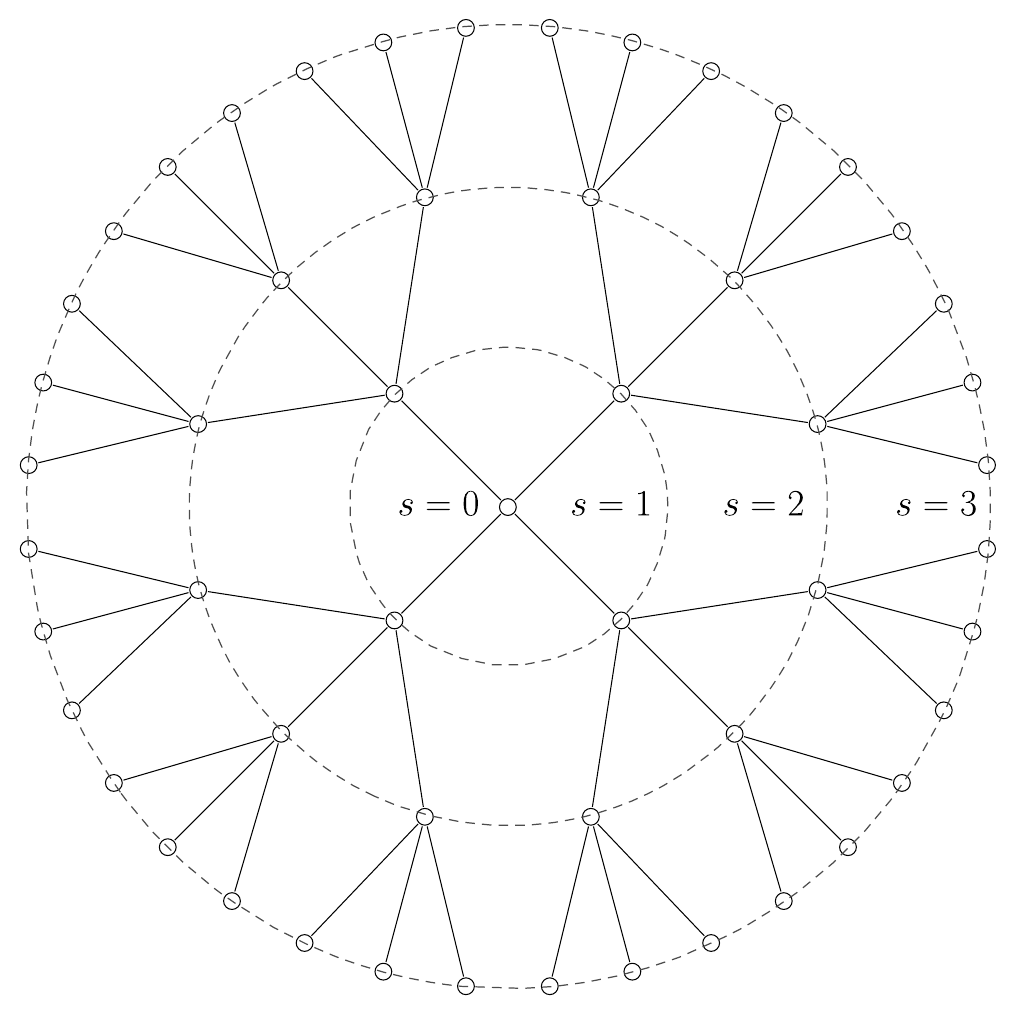}
\caption{Cayley tree with coordination number $q=4$ and three shells ($k=3$). Different sites belonging to the same shell are united by dashed lines. Each shell is labeled with its respective $s$ value.}
\label{fig:cayley}
\end{figure}

A Cayley tree (CT) is a connected graph without loops whose vertices have the same degree (also known as coordination number) $q$, with the exception of those in the boundary of the graph which have only a single neighbor. The simplest way to build a Cayley tree is to start with one site (shell index $s=0$) and add $q$ new neighbors to it to form the first shell ($s=1$) of the tree. From it, one adds $q-1$ new neighbors to all the sites of the previous shell until reaching the last shell of the tree. In Fig.~\ref{fig:cayley} an example of a Cayley tree of coordination number $q=4$ and three shells is shown. The total number of sites in a CT can be calculated as
\begin{equation}
	\label{eq:nt}
	N_T=1 + \frac{q[(q - 1)^k - 1]}{(q - 2)} \hspace{2mm} (q>2),
\end{equation}
where $k$ is the largest value of $s$ for the specific tree ($k=3$ for the tree shown in Fig.\ \ref{fig:cayley}). The number of sites belonging to the $s$-shell is
\begin{equation}
	\label{eq:nk}
	N_s=q(q - 1)^{s - 1} \hspace{2mm} (s>0).
\end{equation}
As the number of shells increases, the ratio between the number of sites belonging to the last shell (the boundary whose sites have only one neighbor) and the total number of sites does not tend to zero as in a regular lattice, but rather converges to $(q-2)/(q-1)$, i.e.\ a macroscopic fraction of sites belong to the boundary, even in the $k\rightarrow\infty$ thermodynamic limit. As a consequence, models formulated on Cayley trees can be dominated by boundary effects. The core of an infinite Cayley tree, whose central part is at an infinite distance of the boundary and thus unaffected by it, is known as the Bethe lattice\cite{Bethe1935,Baxter2016}. Several different systems have been studied using the Bethe lattice as a substrate \cite{Dorogovtsev2008}, like percolation-related models\cite{Chae2012}, diffusion processes \cite{Sahimi1988,Hughes1982}, random aggregates \cite{Krug1988,Bradley1984}, disordered systems~\cite{Angelini2020,Angelini2022}, or transport phenomena \cite{Sahimi1993}.

\subsection{Standard discretization}
\label{sec:ST}

Many previous studies that attempt to integrate the stochastic KPZ equation in finite-dimensional regular lattices use an explicit Euler-Maruyama scheme \cite{Amar1990,Moser1991}. Specifically, in one dimension, the following discretization (with $\Delta x$ lattice spacing) has been frequently used for the Laplacian and the square of the gradient,
\begin{eqnarray}
\label{eq:discretizacion_1d}
\nabla^2 h(x_i,t) &=& \frac{1}{(\Delta x)^2}(h_{i+1} + h_{i-1} - 2h_i), \nonumber \\
(\nabla h)^2(x_i,t) &=& \frac{1}{(2\Delta x)^2}(h_{i+1} - h_{i-1})^2
\end{eqnarray}
so that, using $n$ to denote time steps with time spacing $\Delta t$, the discretized KPZ equation reads \cite{Lam1998,Giada2002}
\begin{equation}
	\label{eq:integracion_euler_1d}
    	h_{i}^{n+1}=h_{i}^n+\frac{\nu\Delta t}{(\Delta x)^2}(h_{i+1}^n+h_{i-1}^n-2h_{i}^n)+ \frac{\lambda\Delta t}{8(\Delta x)^2}(h_{i+1}^n-h_{i-1}^n)^2+\sqrt{2D\Delta t}\hspace{1mm}\eta_i^n\,,
\end{equation}
where $\Delta x=1$ is usually chosen and $\eta_i^n$ is a Gaussian random number with zero mean and unit standard deviation, typically generated with the standard Box-Muller method \cite{Box1958}. In higher dimensions, the discretized equation can be readily obtained by suitably adding terms both to the Laplacian and to the square of the gradient.

In recent years, some developments have been made in the field of partial differential equations defined on discrete networks \cite{Ortega2015,Chung2007}. 
The natural way of extending the definition of the Laplacian and the squared gradient is by taking into account all the neighbors of a given site, i.e.,
\begin{eqnarray}
\label{eq:discretizacion_1d}
\nabla^2 h(x_i,t) &=& \frac{1}{(\Delta x)^2}(h_{i+1} + h_{i-1} - 2h_i), \nonumber \\
(\nabla h)^2(x_i,t) &=& \frac{1}{(2\Delta x)^2}(h_{i+1} - h_{i-1})^2
\end{eqnarray}
where $\mathrm{deg}(i)$ is the degree of the $i$ site (vertex), i.e., the number of neighbors of this site, and the sum $\sum_{j\sim i}$ extends towards all the neighbors $j$ of a given site $i$. All neighbors in the network are considered to be at the same distance to each other, but these expressions can be also generalized to weighted networks \cite{Ortega2015,Chung2007}. The discretized equation writes then
\begin{equation}
	\label{eq:integracion_euler_graf}
    	h_{i}^{n+1}=h_{i}^n+\nu\Delta t\left[\sum_{j\sim i} h_{j}^n-\mathrm{deg}(i)h_{i}^n\right]+ \frac{\lambda\Delta t}{2}\sum_{j\sim i} (h_j^n-h_i^n)^2+\sqrt{2D\Delta t}\hspace{1mm}\eta_i^n\,,
\end{equation}
All the $h$-terms on the right hand side of the equation correspond to the $n$-th time step, i.e., the method is explicit. This way of extending the definitions of the Laplacian and the square of the gradient in its simplest form wil be referred to as the standard (ST) method throughout this paper.

\subsection{Lam and Shin discretization}
\label{sec:LS}

As mentioned earlier, for regular lattices there is another discretization, proposed by Lam and Shin (LS) \cite{Lam1998} for the square of the gradient, in which an additional cross term is added, that makes the equation much more stable numerically. In one dimension, this discretization reads
\begin{equation}
	\label{eq:discretizacion_ls_1d}
    (\nabla h)^2\left(x_i,t\right)=\frac{1}{3(\Delta x)^2} \Big[(h_{i+1}-h_{i})^2+(h_{i}-h_{i-1})^2+(h_{i+1}-h_{i})(h_{i}-h_{i-1})\Big].
\end{equation}
This method can be used for any finite-dimensional lattice by just adding the necessary terms for each additional space dimension. However, there is not an easy extension of this discretization to a network, where directions do not exist and there is not a clear way of picking the pairs to represent the cross term of the LS discretization. The most straightforward approach one might consider is to include all possible pairs in the discretization, namely,
\begin{equation}
(\nabla h)^2\left(\boldsymbol{x},t\right)=\sum_{j\sim i} (h_j-h_i)^2+\sum_{\langle j, k \rangle} (h_j-h_i)(h_i-h_k),
\label{eq:lsn}
\end{equation}
where the sum $\sum_{\langle j, k \rangle}$ in the second term spans over all the pairs of neighbors of site $i$ without repetition; however, Eq.\ \eqref{eq:lsn} \textit{cannot} be used as such, since for $\mathrm{deg}(i)>3$ this expression could lead to a negative value of the square of the gradient. 

With this choice of discretization, the number of quadratic terms grows as $\mathrm{deg}(i)$, while the number of cross terms increases as ${\mathrm{deg}(i) \choose 2} = \mathrm{deg}(i)(\mathrm{deg}(i)-1)/2$. In the LS discretization for a regular lattice, the number of quadratic terms is always twice the number of cross terms. To preserve this ratio and ensure that the discretization of the square of the gradient remains positive-definite, we propose dividing the cross-term summation by $(\mathrm{deg}(i)-1)$. This adjustment maintains the quadratic-to-cross-term ratio at two in all cases. The resulting discretized equation becomes
\begin{eqnarray}\nonumber
	\label{eq:integracion_euler_ls}
    	&&h_{i}^{n+1}=h_{i}^n+\nu\Delta t\left[\sum_{j\sim i} h_{j}^n-\mathrm{deg}(i)h_{i}^n\right]+\sqrt{2D\Delta t}\hspace{1mm}\eta_i^n+
        \\
	    &+&\frac{\lambda\Delta t}{2}\Bigg[\sum_{j\sim i} (h_j^n-h_i^n)^2+\frac{1}{\mathrm{deg}(i)-1}\sum_{\langle j, k \rangle} (h_j^n-h_i^n)(h_i^n-h_k^n)\Bigg]\,.
\end{eqnarray}
In this paper we refer to the integration method associated with Eq.\ \eqref{eq:integracion_euler_ls} as the Lam-Shin (LS) method. For both, the ST and the LS methods, we have used Neumann boundary conditions (BC), i.e., we take the height values of the sites of the last layer to equal those at their parent sites in the penultimate layer. 

\begin{table}[t]
\begin{center}
\begin{tabular}{|c|c|c|c|}
\hline
$q$ & $k$ & $\lambda$  & Number of runs \\ \hline
3  & 4, 6, 8, 10, 12, 14 and 16 & 0.5 & 50 \\ \hline
4 & 4, 6, 8 and 10  & 0.5 & 50 \\
\hline
\multicolumn{4}{c}{}\\
\end{tabular}
\end{center}
\caption{Summary of simulations performed for the KPZ equation using the ST and LS methods. For all these simulations, $\Delta t=0.001$, the number of time steps $N_{\mathrm{steps}}=10^7$, and $D=\nu=1$.}
\label{tab:methods_LS_ST}
\end{table}

\begin{table}[t]
\begin{center}
\begin{tabular}{|c|c|c|c|}
\hline
$q$ & $k$ & $\lambda$  & Number of runs \\ \hline
3  & 4, 6, 8, 10, 12, 14 and 16 & 0 & 50 \\ \hline
4 & 4, 6, 8 and 10  & 0 & 50 \\ \hline
5 & 4, 6 and 8  & 0 & 50 \\ \hline
6 & 4 and 6  & 0 & 50 \\ \hline
\multicolumn{4}{c}{}\\
\end{tabular}
\end{center}
\caption{Summary of simulations performed for the EW equation, where $\lambda=0$ (thus, ST and LS coincide). For all these simulations, $\Delta t=0.001$, the number of time steps $N_{\mathrm{steps}}=10^7$, and $D=\nu=1$.}
\label{tab:EW}
\end{table}

In all our simulations we have fixed $\Delta x=1$. 
For the ST and LS  methods, we have fixed $\Delta t=0.001$ as a good compromise between speed and stability. Nonetheless, we were only able to simulate small values of $\lambda$ due to instability problems. Table \ref{tab:methods_LS_ST} presents a summary of the simulations conducted using the ST and LS methods. Additionally, Table \ref{tab:EW} provides a summary of the conditions under which we simulated the EW equation, where $\lambda = 0$, making irrelevant the choice of integration method for the squared gradient.

\begin{table}[t]
\begin{center}
\begin{tabular}{|c|c|c|c|}
\hline
$q$ & $k$ & $\lambda$  & Number of runs \\ \hline

 3 & 4, 6, \textbf{8}, \textbf{10}, 12, 14 and 16 & 3.0  & 100 \\ \hline
 4 & 4, 6, \textbf{8} and 10 & 3.0  & 100 \\ \hline
 5 & 4, 5, 6, and 7 & 3.0  & 100 \\ \hline
 6 & 4, 5, \textbf{6}, and 7 & 3.0  & 100 \\ \hline
 7 & 4, 5 and 6 & 3.0  & 100 \\ \hline
 8 & \textbf{4}, 5 and 6 & 3.0  & 100 \\ \hline
\end{tabular}
\end{center}
\caption{Summary of simulations performed for the KPZ equation using the controlled stability method. For all these simulations, $\Delta t=0.01$, the number of time steps $N_{\mathrm{steps}}=10^7$, and $D=\nu=1$. The five conditions that appear in bold type were also used to simulate the tensionless KPZ equation in which $\nu=0$.}
\label{tab:methods_control}
\end{table}

\subsection{Controlled instability method using higher powers of the gradient}
\label{sec:CI}

As mentioned above, the method proposed in \cite{Dasgupta1996,Dasgupta1997} is able to control the intrinsic instability of the discretized equation by exchanging the term $(\nabla h_i)^2$ in the discretized equation by $f((\nabla h_i)^2)$, where $f(x)=(1-e^{-cx})/c$, with $c$ being an adjustable parameter. By performing this exchange in Eq.~\eqref{eq:integracion_euler_graf} the thus discretized KPZ equation becomes
\begin{equation}
	\label{eq:integracion_control}
    	h_{i}^{n+1}=h_{i}^n+\nu\Delta t\left[\sum_{j\sim i} h_{j}^n-\mathrm{deg}(i)h_{i}^n\right] + \frac{\lambda\Delta t}{2}f\left(\sum_{j\sim i} (h_j^n-h_i^n)^2\right)+\sqrt{2D\Delta t}\hspace{1mm}\eta_i^n\,.
\end{equation}
Following \cite{Dasgupta1996,Dasgupta1997}, we refer to the integration method associated with Eq.\ \eqref{eq:integracion_control} as the controlled instability (CI) method. The parameter $c$ should be as small as possible to 
accurately approximate the KPZ equation, while being large enough to eliminate numerical instabilities. We found that these appear for values near $c\approx0.001$ or smaller. Although the numerical integration never resulted into overflow for any of the simulations performed, runs using such small $c$ values exhibited clear signs of numerical instability, as, for example, a global roughness that overshoots its saturation value, only to saturate back unpredictably at random times. With this in mind, we set $c=0.01$ for all simulations where this method was applied.

As the use of the CI method efficiently improves the numerical stability of the discretized equation, we have used a larger $\Delta t=0.01$ to reach longer times. Moreover, we did not use Neumann BC as, in contrast with the ST and LS methods, they were not needed to improve the stability of the numerical integration. Hence, when using the CI method, the nodes of the last layer were able to evolve freely following Eq~\eqref{eq:integracion_control}. Table \ref{tab:methods_control} provides a summary of the simulations conducted using this method.

\subsection{Observables}

The main observable that we have measured in this work is the global roughness of the front $w(L,t)$, defined as
\begin{equation}
	\label{eq:width}
	w^2(L,t)=\left\langle \overline{[h_{i}(t)-\bar{h}(t)]^2} \right\rangle,
\end{equation}
where we have used the notation $\overline{(\cdots)} \equiv (1/N_T)\sum_i (\cdots)$ for the space average and $\langle (\cdots) \rangle$ for average over different realizations of the noise. 
Following Oliveira \cite{Oliveira2021}, we have also measured two more roughness-related quantities. Namely, the local roughness, $w_0$, defined as
\begin{equation}
	\label{eq:local_roughness}
	w_0^2=\langle h_0^2\rangle-\langle h_0\rangle^2,
\end{equation}
where $h_0$ is the height value at the central node of the Cayley tree, and the variance, $w^2_{\overline{h}}$, of the average height $\overline{h}(t)$, defined as
\begin{equation}
    \label{eq:mean_fluctuations}
    w_{\overline{h}}^2=\langle \overline{h}^2\rangle-\langle \overline{h}\rangle^2 .
\end{equation}
To further analyze how the surface shape evolves in time, it is interesting to study how the layers grow relative to each other and to the global average of the front. In order to do this, we have measured the difference between the mean heights at the center and at the system boundary \cite{Oliveira2021},
\begin{equation}
	\label{eq:delta_h}
	\Delta \langle h\rangle=|\langle \bar{h} \rangle_0-\langle \bar{h} \rangle_k|,
\end{equation}
where $\langle \bar{h} \rangle_k$ is the mean height value restricted to the outermost ($k$-th) shell or layer, averaged over different noise realizations, and the average growth of the $s$-th layer relative to the global average of the front, i.e.,
\begin{equation}
	\label{eq:capas}
	A(s,t) = \langle \overline{h_i - \bar{h}} \rangle_s ,
\end{equation}
where $s=0,1,\ldots, k$. Note that $\langle  \bar{h} \rangle_0\equiv \langle h_0 \rangle$.

Besides the previous quantities, we have also computed the height-difference correlation function $C_2(r,t)$ relative to the central node of the lattice, namely,
\begin{equation}
	\label{eq:correlation}
	C_2(r,t)=\frac{1}{N_r}\sum_{i\in \mathrm{shell(r)}}{\left\langle \left[h_i(t)-h_0(t)\right]^2\right\rangle},
\end{equation}
where $N_r=q(q - 1)^{r - 1}$ is the number of nodes belonging to the $r$-shell. As the system lacks periodic boundary conditions, this is a natural way of computing the correlations in the tree~\cite{Angelini2020}, 
as being in the $r$-shell is the same as being a distance $r$ away from the central node. This correlation does not seem to have been computed in simulations of KPZ-related models on Cayley trees thus far. Likewise with the last observable that we have measured. Indeed, recent developments on surface kinetic roughening, particularly in the context of KPZ behavior \cite{Kriecherbauer2010,HalpinHealy2015,Takeuchi2018}, have demonstrated that the universal behavior extends beyond the values of the critical exponents for many important universality classes. Specifically, by normalizing the fluctuations of the front around its mean by their time-dependent amplitude as
\begin{equation}
	\label{eq:chi}
	\chi_i(t) = \frac {h_i(t) - \bar{h}(t)}{w(t)} ,
\end{equation}
the probability density function (PDF) of these $\chi$ random variables becomes time-independent within the growth regime and is shared by all members of the universality class \cite{Kriecherbauer2010,HalpinHealy2015,Carrasco2016,Takeuchi2018,Carrasco2019}. In particular, for $d=1$, front fluctuations in the KPZ class follow one of the celebrated Tracy–Widom (TW) distributions depending on the geometry of the system \cite{Kriecherbauer2010,HalpinHealy2015,Takeuchi2018}. While it is unclear what type of distribution should occur on the Bethe lattice, e.g.\ Gaussian behavior might be expected if the effective dimensionality of such a network happened to be larger than the putative upper critical dimension of the model.

The uncertainties of all the observables computed in our simulations have been calculated following the jackknife procedure \cite{Young2015,Efron1982}. Moreover, to perform the averaging of the magnitudes, we define time boxes within which averages are computed and which are evenly spaced in logarithmic time. In all cases, the number of time boxes used was 100. Further details about the use of the jackknife procedure and the time-boxes are provided in Appendix B of Ref.\ \cite{Barreales2020}. 

\section{Results and discussion} \label{sec:results}

In this section we will analyze the evolution of the different observables described in the previous section. First, we will compare the results of the different integration methods proposed for the KPZ equation, also addressing the EW and TKPZ equations as particular cases. After that, we will compare the results of these methods with those of the discrete models reported by Saberi \cite{Saberi_2013} and Oliveira \cite{Oliveira2021} regarding the global and local roughness. Then, we will study the results of the height-difference correlation function and the statistics of front fluctuations. A study of how each layer grows in the Cayley tree for each equation, which will allow us to rationalize our results, is reported in detail in \ref{appendix_A}.

\subsection{Comparison of integration methods}



\begin{figure}[t]
\centering
\includegraphics[width=0.495\textwidth]{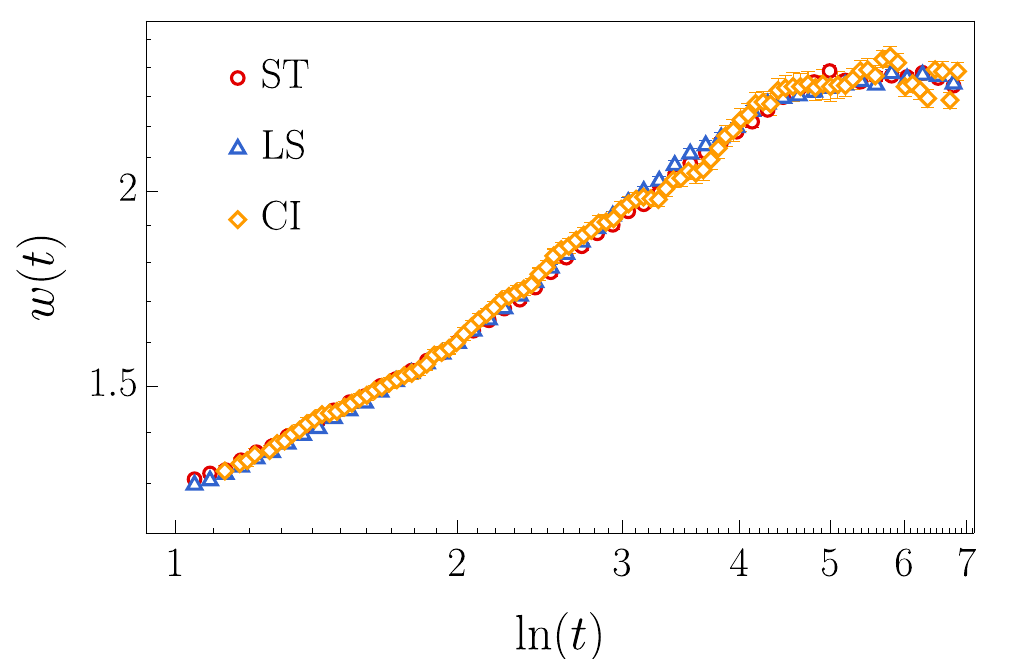}
\includegraphics[width=0.495\textwidth]{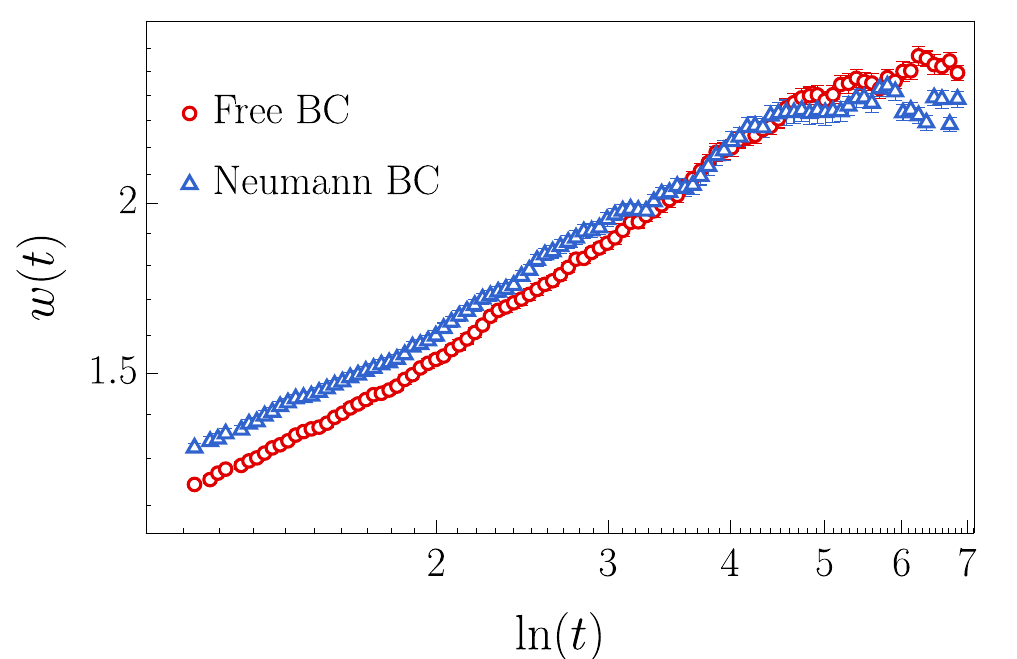}
\caption{Global roughness as a function of time for the KPZ equation. Left panel: Comparison of the three methods introduced in Secs.\ \ref{sec:ST}, \ref{sec:LS}, and \ref{sec:CI}, see legend. 
Right panel: Comparison of different boundary conditions for the CI method, see legend. 
In both panels $q=3$, $k=8$, $\nu=D=1$, and $\lambda=0.5$.}
\label{fig:compara}
\end{figure}

The left panel of Fig.~\ref{fig:compara} shows the behavior of $w(t)$ for the KPZ equation according to the three integration methods introduced in Secs.\ \ref{sec:ST}, \ref{sec:LS}, and \ref{sec:CI}; Neumann BC are employed. The right panel of the same figure shows a comparison between the Free and Neumann BC for the CI integration method. Clearly all three methods yield very similar results within the parameter range where they are all applicable. More importantly, as assessed through the CI method, while changing the BC leads to some quantitative differences, the overall qualitative behavior remains unaffected. This conclusion extends to other observables, like the height-difference correlation function, see \ref{appendix2}.

Although the LS method slightly improves the stability of the integration with respect to ST, in our present case this relative advantage is not as large as in the case of regular lattices. The primary difference between the ST and LS methods, aside from stability, is that the average front value, $\overline{h}$, is lower in the LS method compared to ST. However, both values remain proportional to each other. This outcome is expected as, numerically, the square of the gradient in the LS method is generally smaller than in the ST method, leading to a lower average front value. In the case of the CI method, $\overline{h}$ falls between the values observed in the ST and LS methods while remaining proportional to both. In addition to all this, the ST and LS methods are observed to loose numerical stability with increasing coordination number $q$.

With the ST and LS methods, we have only been able to simulate the KPZ equation for small values of $\lambda$. If we assume that the Cayley tree serves as a suitable substrate for investigating the upper critical dimension of the KPZ class, we are likely operating in the weak-coupling regime, where only the smooth phase of the equation is observable. Neither the ST nor the LS integration methods allow for simulations with larger $\lambda$, and they both break down for the tensionless KPZ equation. This limitation is the primary motivation for our use of the CI method along this paper, although the results of this method are slightly different (but not qualitatively) from those of the other methods. 

In all the discretizations examined, the time increase of the global spatial average of the front, $\bar{h}$, of the front height at the central node of the tree, $h_0$, and of the spatial average of each shell, is linear with $t$ for both, the KPZ and the TKPZ equations, while they are null for the EW equation. Differences in the growth of each layer arise depending on the specific equation studied, which will be analyzed below.

\subsection{Global and local roughness}

\begin{figure}[t]
\centering
\includegraphics[width=0.495\textwidth]{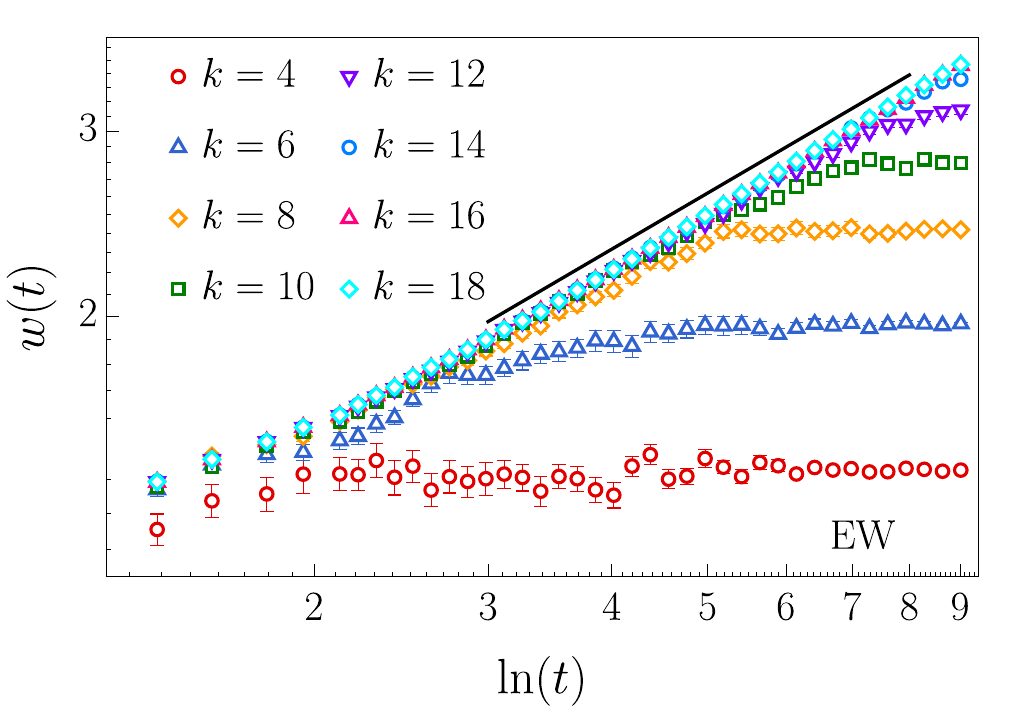}
\includegraphics[width=0.495\textwidth]{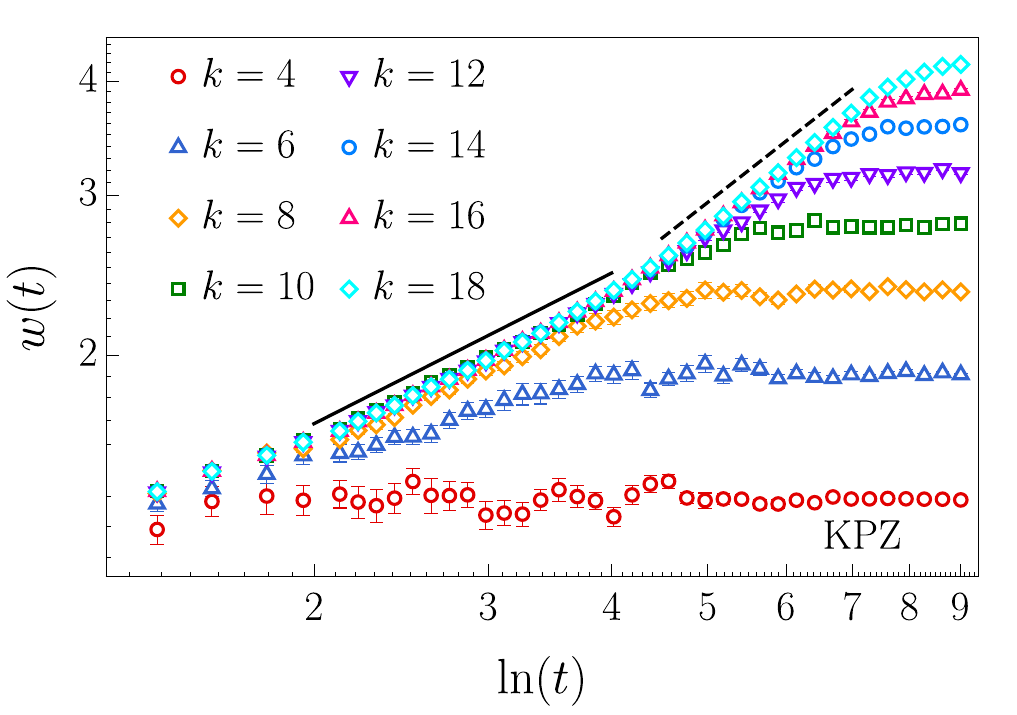}
\caption{Log-log plots of the global roughness $w(t)$ as a function of $\ln t$ for the EW (left panel) and KPZ equations (right panel, where $\lambda=0.5$) on Cayley trees with coordination number $q=3$. Data are provided for different numbers of generations, $k$, of the trees, see legends. As visual references, the solid black lines in both panels correspond to $w\sim (\ln t)^{0.55}$ and the dashed black line in the right panel corresponds to $w\sim (\ln t)^{0.85}$. The integration method used was LS.}
\label{fig:w_integra}
\end{figure}

Figure~\ref{fig:w_integra} shows the evolution of the global roughness $w(t)$ as a function of the logarithm of time for trees of increasing size $k$ for the EW equation (left panel) and for the KPZ equation (right panel).
In Ref.\ \cite{Saberi_2013}, Saberi found a similar logarithmic scaling for several discrete models. Specifically, within the KPZ class he found $w\sim (\ln t)^{0.75}$ for the ballistic deposition (BD) model and $w\sim (\ln t)^{0.57}$ for the restricted solid-on-solid (RSOS) model; he also found $w\sim (\ln t)^{0.51}$ for the random deposition with surface relaxation (RDSR) model, which belongs to the EW class. In our case, the integration of the EW equation clearly yields logarithmic scaling with an exponent close to the one obtained by Saberi. Nonetheless, the same cannot be said about the results obtained from the integration of the KPZ equation since the roughness deviates from this scaling for sufficiently long times, as seen in the right panel of Fig.~\ref{fig:w_integra}. More precisely, the KPZ roughness grows similarly to the EW one during a transient period and then departs from it. This time crossover behavior is familiar for the KPZ equation, e.g.\ in $d=1$, where the roughness grows with different values of the growth exponent for increasing time, namely, $\beta = 1/2$ (as in random deposition) for the shortest times, followed by $\beta = 1/4$ (as in EW) at intermediate times, and finally transitioning to $\beta = 1/3$ (KPZ growth regime) at the longest times prior to saturation to steady state \cite{Forrest1993}.

\begin{figure}[t]
\centering
\includegraphics[width=0.495\textwidth]{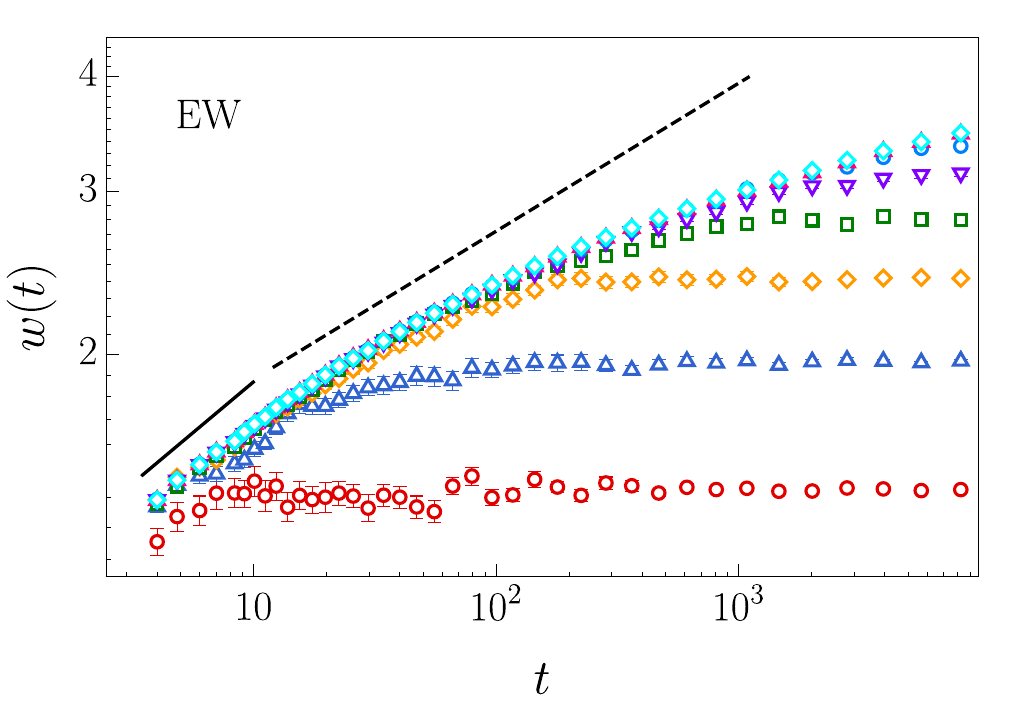}
\includegraphics[width=0.495\textwidth]{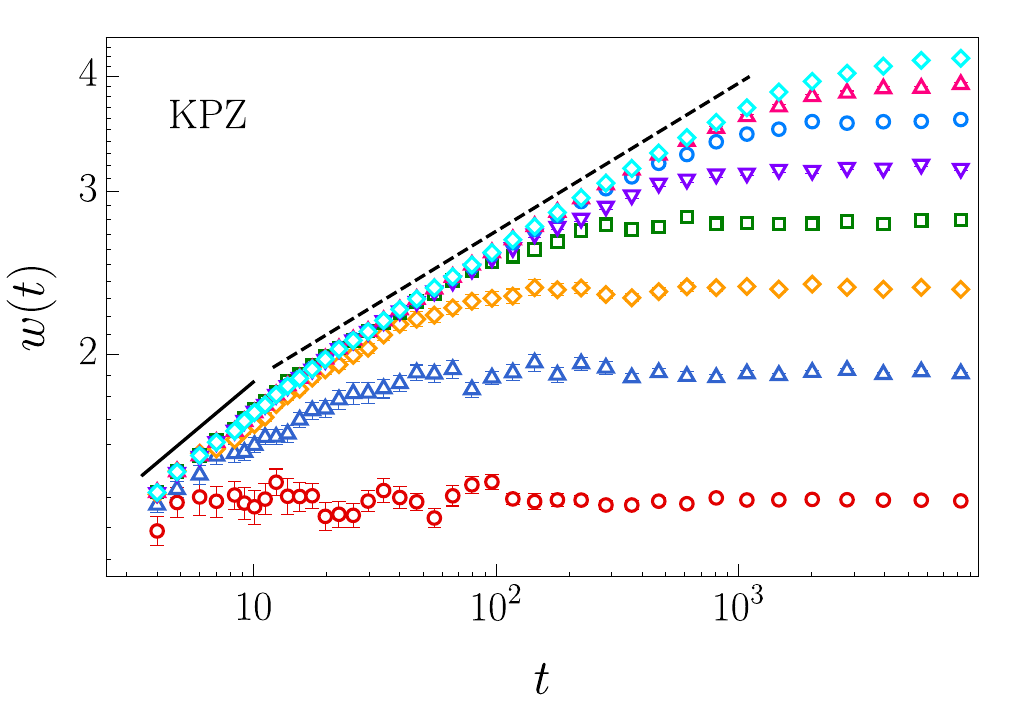}
\caption{Same data as in Fig.\ \ref{fig:w_integra}, but with $t$, rather than $\ln t$, on the horizontal axis. As visual references in both panels, the solid black line now corresponds to $w\sim t^{0.22}$ and the dashed black line corresponds to $w\sim t^{0.16}$.
}
\label{fig:w_noLog}
\end{figure}

Actually, we find that in the KPZ case the growth law for the global roughness $w(t)$ is better captured by a power of $t$, rather than a power of $\ln t$, see the right panel of Fig.~\ref{fig:w_noLog}. In this figure, we show the evolution of the global roughness $w(t)$ as a function of time instead of its logarithm for the EW equation (left panel) and the KPZ equation (right panel). The EW behavior is clearly better described as in Fig.\ \ref{fig:w_integra}. In contrast, for KPZ, while for small $k$ the behavior resembles the EW logarithmic trend, for larger $k>14$ the roughness is better described by a power-law, $w\sim t^\beta$ with $\beta\approx 0.16$, admittedly within a limited time interval prior to saturation. This time crossover behavior seems reminiscent of the EW to KPZ crossover in low dimension just mentioned \cite{Forrest1993}.

\begin{figure}[t]
\centering
\includegraphics[width=0.495\textwidth]{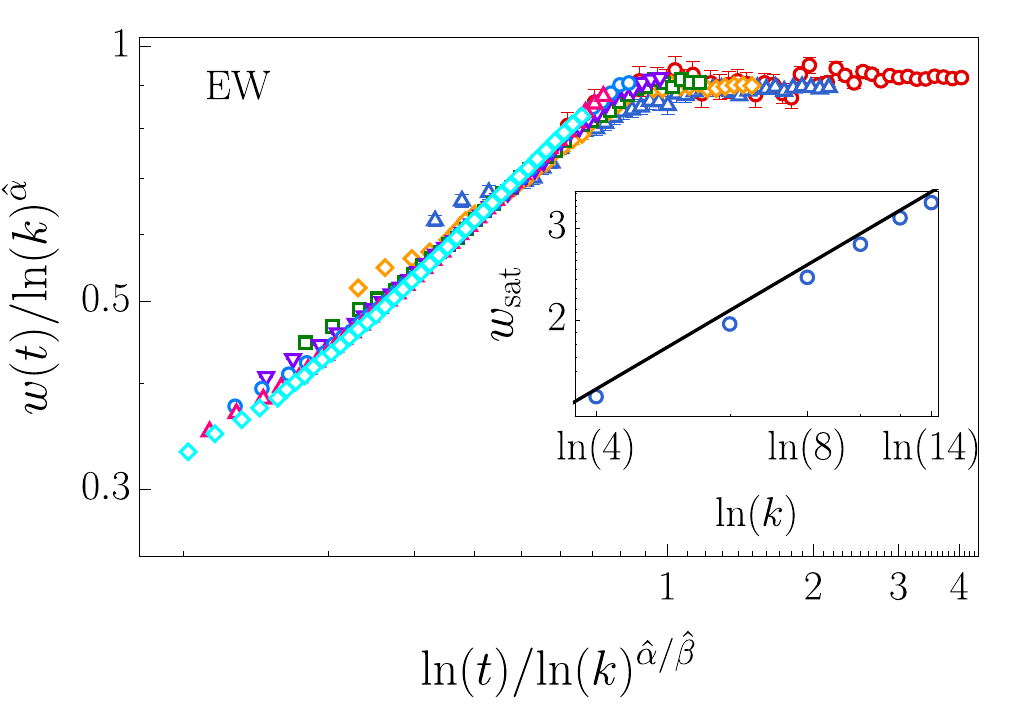}
\includegraphics[width=0.495\textwidth]{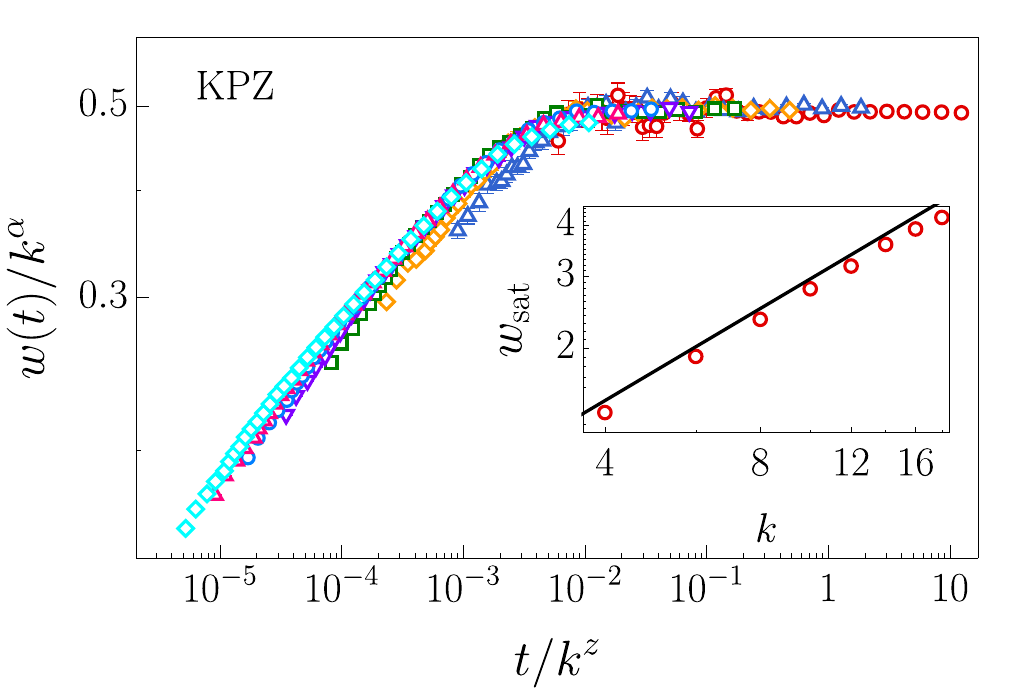}
\caption{Left panel: Data collapse $w(t)/\mathrm{ln}(k)^{\hat{\alpha}}$ vs $\mathrm{ln}(t)/\mathrm{ln}(k)^{\hat{\alpha}/\hat{\beta}}$ for the EW simulations addressed in the left panels of Figs.\ \ref{fig:w_integra} and \ref{fig:w_noLog}, using $\hat{\alpha} = 1.4$ and $\hat{\beta}=0.85$. Inset: Saturation value of the global roughness $w_{\rm sat}$ as a function of the logarithm of the number of layers $\mathrm{ln}(k)$ for the EW equation. As a visual reference, the solid black line corresponds to $w_{\rm sat}\sim \mathrm{ln}(k)^{1.4}$. Right panel: Data collapse $w(t)/k^\alpha$ vs $t/k^z$ for the KPZ simulations addressed in the right panels of Figs.\ \ref{fig:w_integra} and \ref{fig:w_noLog}, using $\alpha = 0.75$ and $z=0.75/0.16=4.69$. Inset: Saturation value of the global roughness $w_{\rm sat}$ as a function of the number of layers $k$ for the KPZ equation. As a visual reference, the solid black line corresponds to $w_{\rm sat}\sim k^{0.75}$.
}
\label{fig:w_colapsos}
\end{figure}

With respect to the system-size dependence of the global roughness, Saberi also reported an analogous logarithmic scaling for the saturation value $w_{\rm sat}\sim (\ln k)^{\hat{\alpha}}$ for the BD and RSOS models, and a more standard scaling $w_{\rm sat}\sim k^\alpha$ for the RDSR model. In our simulations of the EW equation, we have found that the saturation value follows a power law $w_{\rm sat}\sim (\ln k)^{\hat{\alpha}}$, with $\hat{\alpha}\approx1.4$, see the inset in the left panel of Fig.\ \ref{fig:w_colapsos}. Moreover, using the $w(t) \sim (\ln t)^{\hat{\beta}}$ scaling found in the left panel of Fig.\ \ref{fig:w_integra} for $\hat{\beta}=0.55$, the EW data for $w(t)$ approximately collapse onto a single master curve in an analog of the FV data collapse of the global roughness \cite{Barabasi1995}. A similar data collapse was achieved by Saberi for the the BD model. We have not been able to collapse our data in terms of the standard power laws $k^\alpha$ and $k^z$ rather than the $(\ln k)^{\hat{\alpha}}$ and $(\ln k)^{\hat{z}}$ employed for Fig.\ \ref{fig:w_colapsos}. 
The values we obtain for $\hat{\alpha}$ and $\hat{\beta}$ seem to be parameter-dependent, with $\hat{\alpha}$ decreasing when the coordination number $q$ increases, consistent with Saberi's results and in line with expectations that the $d>d_u$ condition is better approximated for increasing $q$. 

In the case of the KPZ equation, 
the saturation value of the global roughness as a function of system size $k$ (see the inset in the right panel of Fig.~\ref{fig:w_colapsos}) is consistent with $w_{\rm sat} \sim k^\alpha$, for $\alpha \approx 0.75$. Combining this with the $w(t) \sim t^{\beta}$ behavior assessed in Fig.\ \ref{fig:w_noLog} for $\beta \approx 0.16$ leads to a proper FV data collapse, see the right panel of Fig.\ \ref{fig:w_colapsos}, using $z=\alpha/\beta \approx 4.69$. 
In this panel, the only deviations in the data collapse occur at short times, where the EW transient is observed.


These results are in good agreement with those reported by Saberi in Ref.\ \cite{Saberi_2013}. However, there are some significant differences. First, one of the most important findings of that reference is the logarithmic scaling of the global roughness for models in the KPZ class. In contrast, we do not observe this scaling unambiguously and the data collapse of the global roughness for the KPZ equation suggests that a standard FV scaling, rather, holds. Moreover, there are some inconsistencies in how the roughness saturation value scales. While we find that, for KPZ, the best scaling is with the number of shells $k$, and for EW, with its logarithm $\ln(k)$, Saberi’s work reports the opposite. However, in our case, and most likely in Saberi’s as well, both scaling behaviors may be compatible due to the limited range of the number of shells analyzed. Moreover, the behavior we observe as the coordination number $q$ increases differs from Saberi's findings. While Saberi, simulating the BD model, reported a higher saturation value for larger $q$, our integration of the KPZ equation shows a decreasing saturation value as $q$ increases. Moreover, we found some surprising cases where the saturation value did not increase with the coordination number. For instance, for $ k=6 $, the value for $ q=7 $ is larger than for $ q=6 $, and close to that for $ q=5 $.


In summary, the result obtained for the EW equation, namely the scaling behavior of the global roughness $w(t)$ with the logarithm of time, reveals that finite Cayley trees are not a suitable substrate for investigating the infinite-dimensional limit of this type of stochastic equations, since for EW the upper critical dimension, where this kind of scaling occurs, is $d_u^{\mathrm{EW}} = 2$, as mentioned in the introduction. A more detailed discussion of the relationship between the global roughness, the local roughness $w_0$, and the variance of the mean height $w_{\overline{h}}$, along with results for the TKPZ equation, is given in \ref{appendix_0}.

\subsection{Height-difference correlation function}

\begin{figure}[t]
\centering
\includegraphics[width=0.495\textwidth]{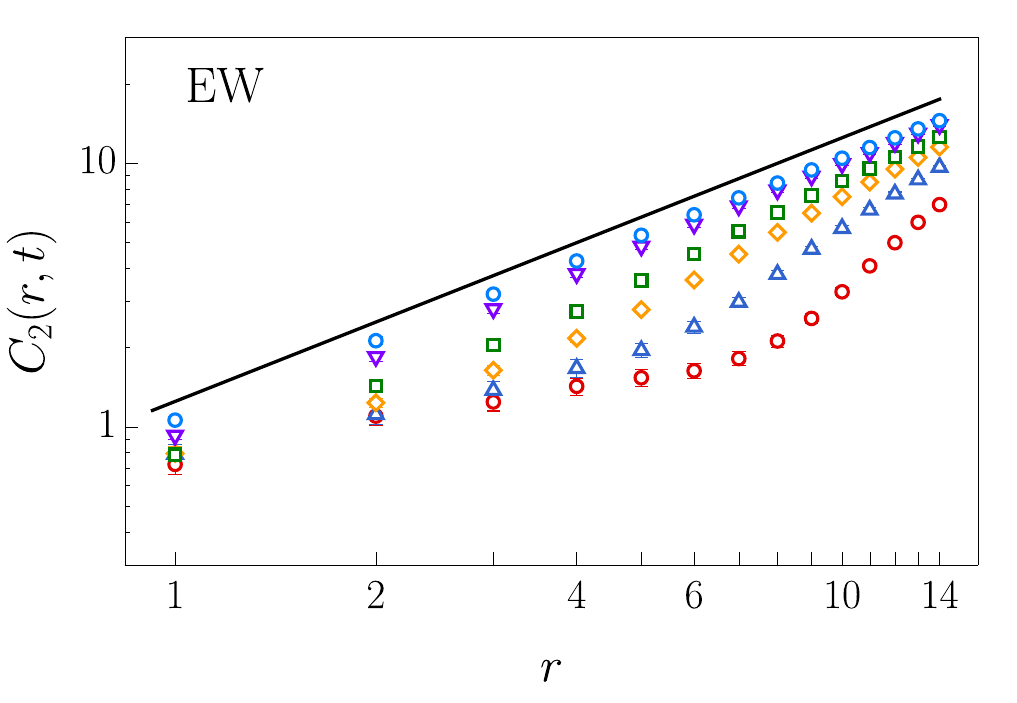}
\includegraphics[width=0.495\textwidth]{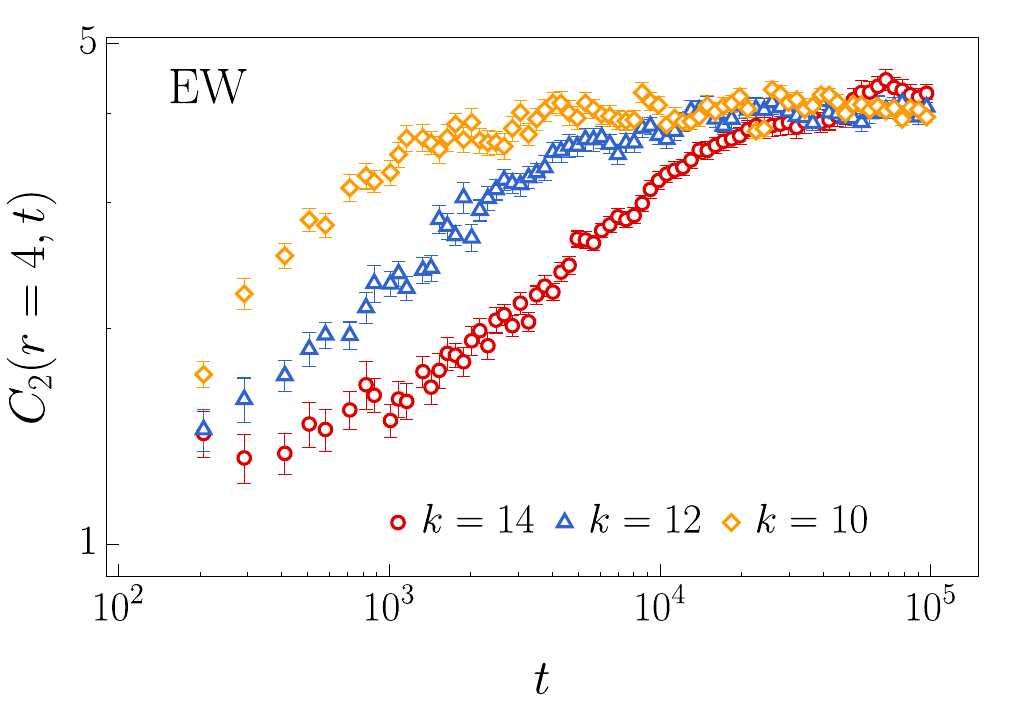}
\caption{Left panel: Height-difference correlation function $C_2(r,t)$ as a function of $r$ for $t=10,30,50,60,80$, and $100$, bottom to top, for the EW equation using $q=3$ and $k=14$. As a visual reference, the solid black line corresponds to $C_2(r,t)\sim r$. Right panel: Time evolution of the height-difference correlation function $C_2(r,t)$ for fixed $r=4$, $q=3$, and three system sizes $k$, see the legend.}
\label{fig:C2_EW}
\end{figure}

We continue by assessing the behavior of real-space correlation functions, which have scarcely been addressed earlier for surface growth models on Cayley trees. Figure~\ref{fig:C2_EW} shows the results of the height-difference correlation function for the EW equation. The left panel displays $ C_2(r,t) $ as a function of the distance to the center, $ r $, at various times, where larger values of $ C_2(r,t) $ correspond to longer times. 

From the point of view of kinetic roughening and according to the simplest Family-Vicsek dynamic scaling ansatz below the upper critical dimension, $C_2(r,t)$ is expected to scale as \cite{Barabasi1995,Krug1997}
\begin{equation}  
    \label{eq:correlation_length_loc}
    C_2(r,t) \sim \left\{ \begin{array}{l}
    r^{2\alpha}\; \mbox{if} \; r\ll\xi(t), \\
    t^{2\beta}\;  \mbox{if} \; r\gg\xi(t), \end{array}
    \right.
\end{equation}
where $\xi(t) \sim t^{1/z}$ is the correlation length. If we assume that the system has saturated [thus, $ r \ll \xi(t) $] for the longest times shown in the left panel of Fig.\ \ref{fig:C2_EW}, this would imply $ \alpha = 1/2 $, which matches the roughness exponent value for the EW class in $ d = 1 $. This exponent is consistent across all our simulations for the EW equation.

Note, however, that the detailed time evolution of the correlation data do not fully comply with the expectation, Eq.\ \eqref{eq:correlation_length_loc}, for short times. Specifically, the right panel of Fig.\ \ref{fig:C2_EW} illustrates the time evolution of the height-difference correlation function at a fixed distance from the center ($ r=4 $). In this panel, the correlation function at a given distance can be observed to reach a saturation value which appears to be largely independent of the system size $ k $. Additionally, the time at which the correlation function saturates coincides with the saturation time of the global roughness $ w(t) $ (expected to scale as $k^z$) while, according to Eq.\ \eqref{eq:correlation_length_loc}, $C_2(r,t)$ is expected to saturate at the scale-dependent time $r^z$. From this point of view, the time-dependent behavior the numerical data for $C_2(r,t)$ seems reminiscent of so-called anomalous surface kinetic roughening \cite{Krug1997,Lopez1997,Ramasco2000}.

In our present system, the behavior observed in Fig.~\ref{fig:C2_EW} can be explained as follows. Once the front has saturated, and given that correlations are measured from the center, each branch of the network, from the central node to the outermost layer, acts as a one-dimensional chain with free edges. 
Since Cayley trees have no loops, we argue that the presence of branches does not influence the correlation function. This idea will be further reinforced in the next section, where the fluctuation distribution will reveal the Gaussian behavior characteristic of the EW class. Additional examples of this behavior exist; for instance, the correlations in the Ising model on the Cayley tree are known to behave as in the one-dimensional system \cite{Dorogovtsev2008}, and likewise for bond percolation \cite{Christensen2005}. The result from the study of the correlations of the EW equation, that is, correlations similar to those in $d=1$, reinforces the finding from the previous section, where we had already pointed out that Cayley trees do not appear to be a suitable substrate for studying the infinite-dimensional limit of these equations.
\begin{figure}[t]
\centering
\includegraphics[width=0.495\textwidth]{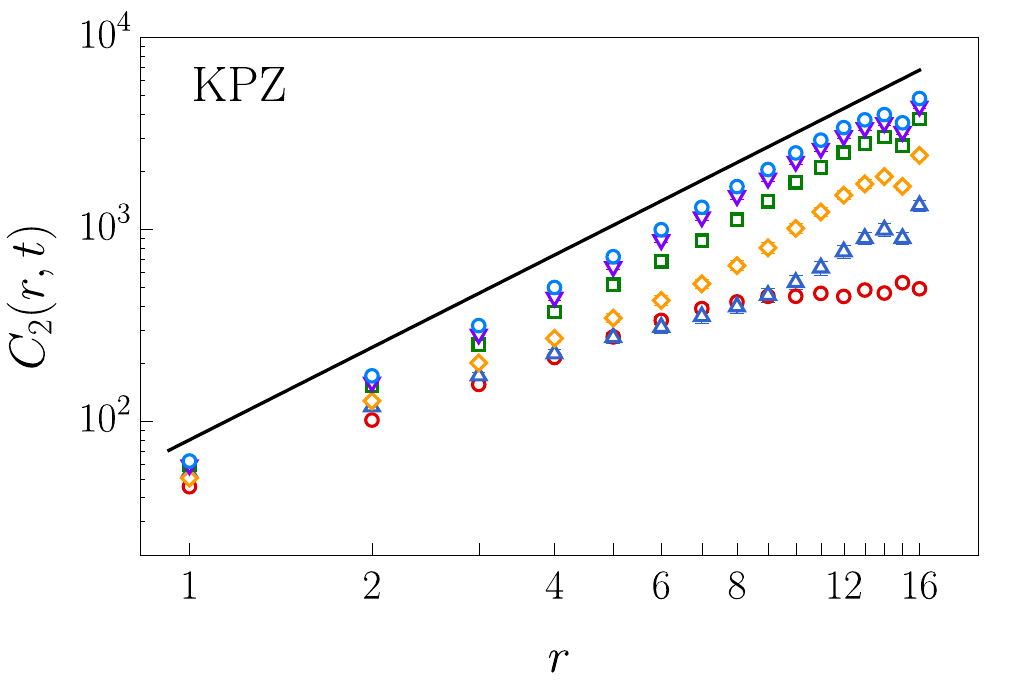}
\includegraphics[width=0.495\textwidth]{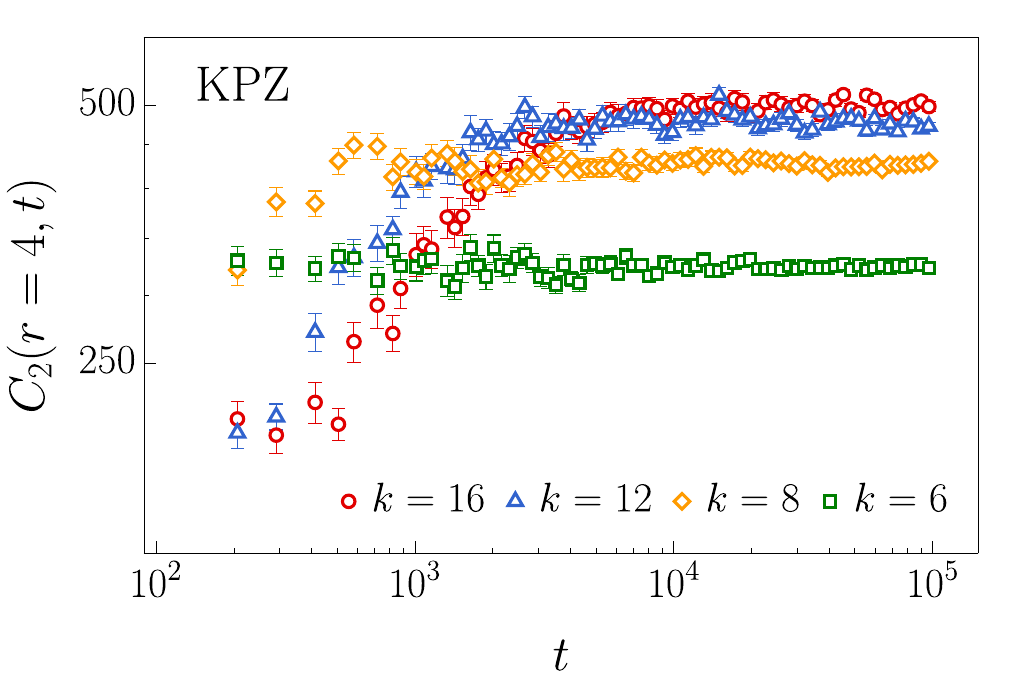}
\caption{Left panel: Height-difference correlation function $C_2(r,t)$ as a function of $r$ for $t=10,20,30,40,50$, and $100$, bottom to top, for the KPZ equation using $q=3$, $k=16$, and $\lambda=3$. As a visual reference, the solid black line corresponds to $C_2(r,t)\sim r^{1.6}$. Right panel: Time evolution of the height-difference correlation function $C_2(r,t)$ for fixed $r=4$, $q=3$, and for various system sizes $k$, see the legend. The integration method used was CI.}
\label{fig:C2_WC}
\end{figure}

We next consider the behavior of the height-difference correlation function for the KPZ equation, see Fig.~\ref{fig:C2_WC}. The left panel displays $ C_2(r,t) $ as a function of the distance to the center, $ r $, at various times, where larger values of $ C_2(r,t) $ correspond to later times. The $C_2(r,t) \sim r^{1.6}$ behavior reached for the longest times would correspond to $\alpha\approx 0.8$, not far from the $\alpha\approx 0.75$ value implied by the data collapse of the global roughness data shown in Fig.\ \ref{fig:w_colapsos} for KPZ. For the sake of reference, recall that $\alpha=1/2$ for 1D KPZ, see Eq.\ \eqref{eq:exponentes_kpz_1d}. 

The right panel in Fig.\ \ref{fig:C2_WC} illustrates the time evolution of the height-difference correlation function at a fixed distance from the center ($ r=4 $). In this panel, it is evident that the correlation function rapidly saturates after a short transient. Moreover, the saturation value at a fixed distance increases with the system size $ k $. As for the EW equation, the saturation time of the correlation function coincides with the saturation time of the global roughness $ w(t) $.

While the shape of the correlation function shown in the left panel of Fig.~\ref{fig:C2_WC} is relatively similar to that obtained for the EW equation, there are a few substantial differences. The first one is that the effective roughness exponent changes for each condition studied, being always larger than one. Hence, for no parameter conditions does $C_2(r,t)$ reproduce the 1D KPZ behavior. Moreover, such an effective roughness exponent value grows with both, the coordination number $q$ and the number of layers $k$. This can be explained by considering that the nonlinear term in the KPZ is playing a role in the correlations. As $q$ and $ k $ increase, the influence of external branches becomes more significant in the dynamics, further deviating the behavior of the KPZ equation from that of the EW equation. The second difference is the presence of a jump in the penultimate layer for the KPZ equation, with a value which is consistently smaller than expected. This behavior is discussed in detail in \ref{appendix_A}. 

\begin{figure}[t]
\centering
\includegraphics[width=0.495\textwidth]{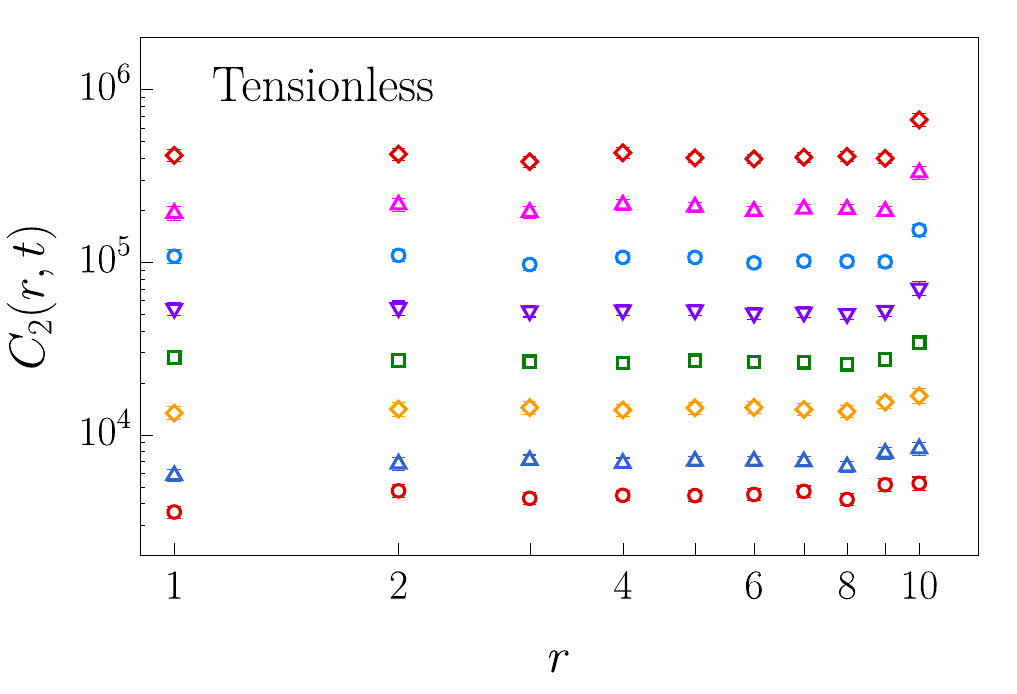}
\includegraphics[width=0.495\textwidth]{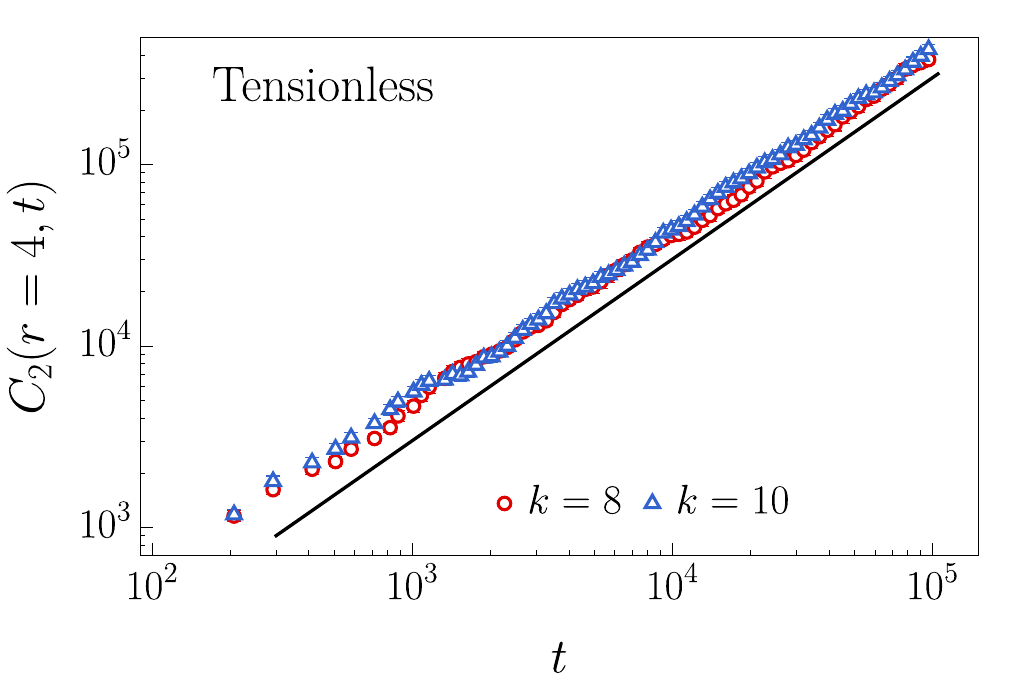}
\caption{Left panel: Height-difference correlation function $C_2(r,t)$ as a function of $r$ for $t=30,40,50,60,70,80,90$, and $100$, bottom to top, for the tensionless KPZ equation using $q=3$ and $k=10$. Right panel: Time evolution of the height-difference correlation function $C_2(r,t)$ for fixed $r=4$, $q=3$, and two system sizes, see the legend. As a visual reference, the solid black line corresponds to $C_2(r=4,t)\sim t$. The integration method used was CI.}
\label{fig:c2_tensionless}
\end{figure}

Finally, Fig.~\ref{fig:c2_tensionless} shows the results of the height-difference correlation function for the tensionless KPZ equation. The left panel displays $ C_2(r,t) $ as a function of the distance to the center, $ r $, at various times, where larger values of $ C_2(r,t) $ correspond to later times. The jump present in the last layer is also discussed in \ref{appendix_A}. The right panel illustrates the time evolution of the height-difference correlation function at a fixed distance from the center ($ r=4 $). In this panel, the correlation function is observed to grow indefinitely, never reaching saturation. In fact, it grows at the same rate as the (squared) global roughness for this equation (recall Fig.\ \ref{fig:compara_3w}), $C_2(r,t)\sim t \sim w^2(t)$, as expected for random deposition \cite{Barabasi1995}. The (uncorrelated) $r$-independent shape of $C_2(r,t)$ shown in the left panel of Fig.~\ref{fig:c2_tensionless} strengthens this interpretation in terms of the RD model, which similarly lacks correlations and never reaches saturation. Indeed, we have also measured the height-difference correlation function in our simulations of the RD equation on CT, see \ref{appendix3}. The results are shown on Fig.\ \ref{fig:RD2}, which is virtually identical to Fig.\ \ref{fig:c2_tensionless} for the TKPZ equation, except for the jump in the last value of $C_2(r,t)$ which occurs for the largest $r$ and all values of time, as just mentioned for TKPZ. For reference, for $d=1$ the TKPZ height-difference correlation function scales non-trivially with $r$, with local roughness exponent value $\alpha_{\rm loc}=1/2$ \cite{RodriguezFernandez2022}. Intriguingly, for the \textit{derivative of} the 1D TKPZ equation (the stochastic inviscid Burgers equation), $C_2(r,t)$ does behave as in the left panel of Fig.\ \ref{fig:c2_tensionless}, the only difference being that saturation to steady state is (perhaps surprisingly) achieved in that case \cite{RodriguezFernandez2022}.


\subsection{Statistics of height fluctuations}

\begin{figure}[t]
\centering
\includegraphics[width=0.495\textwidth]{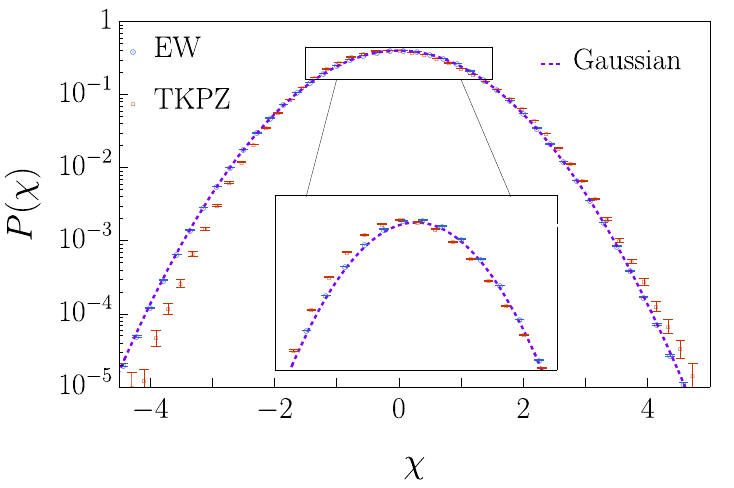}
\includegraphics[width=0.495\textwidth]{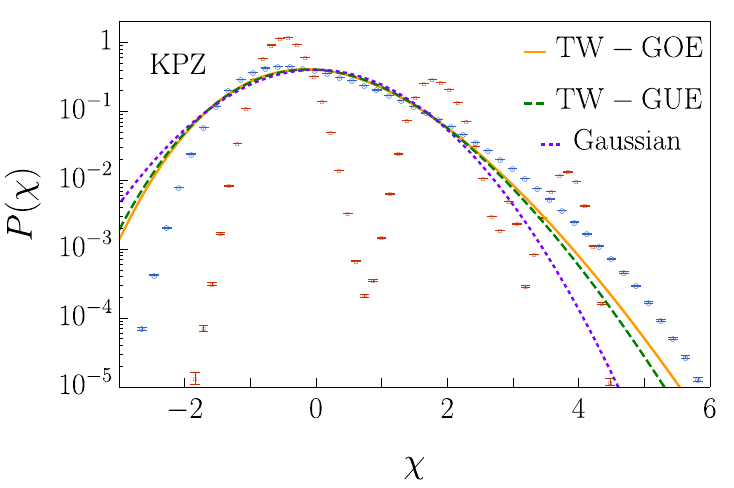}
\caption{
Fluctuation histograms of the rescaled height fluctuations, $\chi$, see Eq.~\eqref{eq:chi}. The left panel shows results for the EW and the TKPZ equations, see legend, for $q=3$ and $k=10$. The inset shows a zoom of the boxed area for the central part of the distributions in the $-1.5 < \chi < 1.5$ interval. The right panel shows two conditions for the KPZ equation with $\lambda=3$, namely $k=10$ for $q=3$ (blue circles) and $k=4$ for $q=6$ (red squares). In both panels, the dotted purple line corresponds to a Gaussian distribution. In the right panel the solid orange and the green dashed lines correspond to the GOE and GUE Tracy-Widom distributions, respectively. The integration method used was CI.}
\label{fig:chi}
\end{figure}

We close our numerical study of various traits of kinetic roughening universality classes for the EW, KPZ, and TKPZ equations on Cayley trees with a study of height fluctuations. Specifically, Fig.~\ref{fig:chi} presents histograms of rescaled height fluctuations $\chi$, Eq.\ \eqref{eq:chi}, for the EW and the tensionless KPZ equations on the left panel, and for two cases ($q=3$, $k=10$ and $q=6$, $k=4$) of the KPZ equation on the right panel. For reference, the figure also shows the exact forms of the Gaussian distribution and of the Tracy-Widom distributions corresponding to the Gaussian orthogonal (TW-GOE) and Gaussian unitary (TW-GUE) ensembles. The fluctuation PDF for the EW class for any $d$ below $d_u$ is known to be Gaussian \cite{Barabasi1995,Krug1988,Prolhac2011}. For the 1D KPZ equation, the PDF in the nonlinear growth regime is provided by TW-GOE for e.g.\ band geometry using periodic BC \cite{Takeuchi2018} and TW-GUE for an interface with an overall circular geometry \cite{Takeuchi2018,Santalla2014}. Note, for the 1D TKPZ equation with periodic BC, the fluctuation PDF, albeit being non-Gaussian, is known not to be TW-GOE \cite{RodriguezFernandez2022}.

The left panel of Fig.\ \ref{fig:chi} cleraly confirms a Gaussian PDF for our simulations of the EW equation on the CT. This behavior remains consistent across all our simulations of this continuum model. On the other hand, the PDF of height fluctuations for the tensionless KPZ equation is clearly non-Gaussian (in particular, it is not even in $\chi$), see further below.
While there are no particular expectations on, e.g., TW behavior for the TKPZ equation \cite{RodriguezFernandez2022}, the present comparison underscores the non-Gaussian behavior of the PDF we presently obtain for this model on the CT.
Hence, the TKPZ equation departs in this aspect from random deposition, for which the fluctuation distribution is Gaussian, see Fig.\ \ref{fig:RD1} in \ref{appendix3}. Note, in the context of the KPZ equation, the non-zero skewness (asymmetry) of the fluctuation PDF is a landmark of the nonlinear term \cite{Takeuchi2018}. Overall, the PDF behavior of the TKPZ equation could thus be a strong indication that the observed trends of the global roughness $ w(t) $ and height-difference correlation function are indeed nontrivial even while heavily influenced by boundary effects, see \ref{appendix_A}. 

In the right panel of Fig.~\ref{fig:chi}, two conditions are presented for the KPZ equation. None follows ananlytically known distributions like Gaussian or TW. Furthermore, the behavior observed for different KPZ equation conditions varies from case to case. In some instances, such as one of the conditions shown in the right panel of Fig.~\ref{fig:chi}, oscillations appear. These oscillations can be explained by analyzing the growth dynamics of each layer in the KPZ equation, see ~\ref{appendix_A}.

The interpretation of our numerical results for the fluctuation PDF of the various equations are further supported by explicit results on the cumulants of these height distributions. Thus, we have additionally evaluated the third and fourth order cumulants, namely the skewness and excess kurtosis, whose values are analytically known for the Gaussian, TW-GOE, and TW-GUE distributions appearing in Fig.\ \ref{fig:chi} \cite{Kriecherbauer2010,Takeuchi2011}. Writing the local height fluctuations as $\delta h_i(t)=h_i(t)-\bar{h}(t)$, we define the skewness and the excess kurtosis as $S~=~\langle \delta h_i(t)^3 \rangle / \langle \delta h_i(t)^2 \rangle^{3/2}$ and $K=\langle \delta h_i(t)^4 \rangle / \langle \delta h_i(t)^2 \rangle^{2}-3$, respectively. Specifically, for the cases shown in Fig.~\ref{fig:chi}, we have obtained $S=0.00(6)$ and $K=0.08(1)$ for the EW case, while $S=0.172(4)$ and $K=0.055(7)$ for the tensionless KPZ case. For the KPZ equation, the values are $S = 0.74(1)$ and $K = 0.52(2)$ for the $q = 3$, $k = 10$ condition, while for $q = 6$ and $k = 4$, the values we obtain are $S = 1.471(1)$ and $K = 1.022(1)$, respectively. \footnote{For comparison, the precise skewness and excess kurtosis values are $S=0.29346452408$ and $K=0.1652429384$ for the TW-GOE and $S=0.224084203610$ and $K=0.0934480876$ for the TW-GUE \cite{Bornemann2010}.}

Table~\ref{table:results} provides a summary of the results obtained for the three equations that have been studied on CT in this work, namely, EW, KPZ, and tensionless KPZ.

\begin{table}[t]
    \centering
    \renewcommand{\arraystretch}{1.5}
    \begin{tabular}{|c|c|c|c|}
        \hline
               & \textbf{EW} & \textbf{KPZ} & \textbf{TKPZ} \\
        \hline
        \hline
        $\langle h \rangle_k$ & $\langle h \rangle_k \approx 0$ & $\langle h \rangle_k \sim t$ & $\langle h \rangle_k \sim t$ \\
        \hline
        $\langle\bar{h}\rangle$ & $\langle\bar{h}\rangle= 0$ & $\langle\bar{h}\rangle \sim t$ & $\langle\bar{h}\rangle \sim t$ \\
        \hline
        $w(t)$ & \begin{tabular}{@{}c@{}} $w \sim (\ln t)^{\hat{\beta}}$ \\ $\hat{\beta}$ depends on $q$ and $k$ \end{tabular} & \begin{tabular}{@{}c@{}} $w \sim (\ln t)^{\hat{\beta}}$ (EW transient) \\ $w \sim t^{\beta}$ (prior to saturation) \\ $\hat{\beta}$ and $\beta$ depend on $q$ and $k$ \end{tabular} & \begin{tabular}{@{}c@{}} $w \sim t^{\beta}$ \\ $\beta=1/2$ \end{tabular}\\
        \hline
        $w_{\rm sat}$ & \begin{tabular}{@{}c@{}} $w_{\rm sat} \sim (\ln k)^{\hat{\alpha}}$ \\ $\hat{\alpha}$ depends on $q$ \end{tabular} & \begin{tabular}{@{}c@{}} $w_{\rm sat} \sim k^{\alpha}$ \\ $\alpha$ depends on $q$ 
        \end{tabular} & \begin{tabular}{@{}c@{}} No saturation \end{tabular}\\
        \hline
        $w_0(t)$ & $w_0 \sim t^{1/2}$ & $w_0 \sim t^{1/2}$ & $w_0 \sim t^{1/2}$ \\
        \hline
        $w_{\bar{h}}(t)$ & $w_{\bar{h}} \sim t^{1/2}$ & $w_{\bar{h}} \sim t^{1/2}$ & $w_{\bar{h}} \sim t^{1/2}$ \\
        \hline
        $C_2(r,t)$ & \begin{tabular}{@{}c@{}} $C_2(r,t) \to r$ \end{tabular} & \begin{tabular}{@{}c@{}} $C_2(r,t) \to r^{2\alpha}$ \\ Jump in penultimate layer \end{tabular} & \begin{tabular}{@{}c@{}} $C_2(r,t) \sim \mathrm{const.}$ \\ Jump in last layer \end{tabular} \\
        \hline
        $P(\chi)$ & \begin{tabular}{@{}c@{}} Gaussian 
        \end{tabular} & \begin{tabular}{@{}c@{}} No clear shape \\ Oscillations for some conditions \end{tabular} & \begin{tabular}{@{}c@{}} Non-Gaussian 
        \end{tabular} \\
        \hline
        $\Delta \langle h\rangle$ & \begin{tabular}{@{}c@{}} $\Delta \langle h\rangle \approx 0$ \\ for all layers \end{tabular} & \begin{tabular}{@{}c@{}} $\Delta \langle h\rangle$ saturates \\ $\Delta \langle h\rangle_\infty \sim k^{\lambda}$ \end{tabular} & $\Delta \langle h\rangle \sim t^{\delta}$ \\
        \hline
        $A(s,t)$ & \begin{tabular}{@{}c@{}} $A(s,t) \approx 0$ \\ for all layers \end{tabular} & \begin{tabular}{@{}c@{}} Last layer slower than $\bar{h}$ \\ Other layers faster than $\bar{h}$ 
        \\ Stationary value reached \\ Stratified; 
        causes $P(\chi)$ oscillations  \end{tabular} & \begin{tabular}{@{}c@{}} Last layer slower than $\bar{h}$ \\ Other layers faster than $\bar{h}$ \\ No stationary value 
        \end{tabular} \\
        
        \hline
    \end{tabular}
    \caption{Summary of results for the three stochastic partial differential equations that have been studied on CT in this work.}
    \label{table:results}
\end{table}

\section{Conclusions}\label{sec:concl}

In this work, we have numerically integrated the Kardar-Parisi-Zhang (KPZ) equation on Cayley trees, analyzing in depth three three distinct cases, namely, the EW, the standard KPZ, and the tensionless KPZ equations. Additionally, in \ref{appendix3} we present results for the Random Deposition (RD) equation. 


Specifically, we have compared several discretization methods, assessing both their numerical stability and their ability to capture the growth dynamics accurately. Our analysis revealed that the ST and LS schemes effectively reproduce the behavior expected from discrete models in the KPZ class; however, they exhibit numerical instabilities for large values of the parameter $\lambda$ of the nonlinear term. In contrast, the CI method proved essential for stabilizing the numerical integration under these conditions, allowing us to explore a broader region of parameter space, as well as the TKPZ equation. This method also supports longer simulation times, which are crucial for studying the behavior of the system at long times. The results obtained with the three methods were largely indistinguishable. Additionally, we investigated the dependence of the main observables on the boundary conditions. While some differences were observed between Free and Neumann BC, the principal results and conclusions of this study remain largely unaffected by the choice of BC.

Our results closely replicate those of previous simulations on Bethe lattices of discrete growth models in the KPZ and EW universality classes. For the EW equation, the global roughness exhibits logarithmic scaling, in agreement with the findings reported by Saberi~\cite{Saberi_2013}. In contrast, the KPZ equation shows a more intricate behavior: the global roughness initially grows in a manner similar to the EW case during a short transient regime, but then transitions to a distinct growth phase characterized by nontrivial power-law scaling. A detailed analysis of this later regime is challenging, as the system rapidly reaches saturation. Remarkably, in the case of the tensionless KPZ equation, the roughness increases indefinitely, following a growth pattern reminiscent of that observed in the RD equation.

We have further investigated the local height fluctuations by analyzing the behavior of the local (squared) roughness $w^2_0(t)$ and the variance of the average height, $w^2_{\bar{h}}(t)$. Under conditions such that the global roughness has saturated for the EW and KPZ equations, $w^2_{0}(t)$ and $w^2_{\bar{h}}(t)$ continue increasing indefinitely with time for long enough times, as a consequence of the models being non-deterministic \cite{Oliveira2021}. In contrast, for the RD and tensionless KPZ equations, all three roughness functions increase linear with $t$ indefinitely, with the variance of the average height being smaller than the other two quantities. Hence, for these equations, the $w(t) \sim t^{1/2}$ growth of the global roughness is intrinsic, and not merely a consequence of the growth of $w^2_{\bar{h}}(t)$.

Additionally, we have also computed real-space correlation functions, specifically the height difference correlation function $C_2(r,t)$. For the EW and KPZ equations, at sufficiently long times we obtain power-law behavior with effective roughness exponent values which, respectively, compare well with those found in $d=1$ or with that (substantially large one) obtained from the corresponding $w(t)$ data. Meanwhile, the tensionless KPZ equation yields trivial space correlations, virtually identical (except for some features) to those obtained and expected for the RD equation.

Finally, regarding fluctuation distributions, our results confirm that height fluctuations in the EW and RD equations on CT follow a Gaussian distribution, as expected. For the KPZ equation, the probability distribution function depends on system conditions; in certain cases, oscillatory features emerge due to the relative growth dynamics between layers. In the tensionless KPZ case, the fluctuation statistics exhibit some similarities to the Tracy–Widom distribution, though significant deviations appear at higher coordination numbers while keeping a non-zero skewness. These observations indicate that fluctuation behavior in growth equations featuring a KPZ nonlinearity on tree-like structures differ markedly from what is obtained on regular lattices.

Overall, the combination of all these findings question the suitability of the Bethe lattice, or more precisely, finite Cayley trees, as a straightforward infinite-dimensional limit of hypercubic lattices for the EW, KPZ, or tensionless KPZ stochastic partial differential equations. Instead, they exhibit pronounced finite-size and boundary effects. For instance, for the EW equation, if the Bethe lattice truly approximated the infinite-dimensional limit, the interface would remain smooth, given that the upper critical dimension for the EW equation is $d_u^{\mathrm{EW}} = 2$.

An important aspect of our analysis is the influence of boundary effects on the growth process. Due to the distinctive structure of Cayley trees, the outermost layer grows more slowly than the average interface when the nonlinear term is present (see \ref{appendix_A}), since each node in this layer is connected to only one neighbor. In contrast, the central node experiences the fastest growth. This asymmetry propagates through the intermediate layers, leading to nontrivial correlations and deviations from standard scaling behavior. The analysis of the saturation of height differences between the center and the boundary shows that KPZ surfaces on Cayley trees remain macroscopically curved in the thermodynamic limit ($k \rightarrow \infty$). This result supports Oliveira’s conclusion that boundary effects hinder reliable measurements of global roughness~\cite{Oliveira2021}. Furthermore, in the case of the tensionless KPZ equation, the height differences between successive layers do not fully saturate.

While the numerical integration methods we have employed do seem to provide a robust framework for analyzing growth dynamics on networked structures, the strong influence of boundary effects necessitates caution in the interpretation of results. Future research could explore alternative network topologies that better emulate high-dimensional behavior while minimizing artifacts introduced by boundary conditions. One possible direction would be to integrate this type of equation into networks \cite{Dorogovtsev2008} such as the Watts-Strogatz model, in order to observe differences as the network loses regularity when the rewiring probability of the model increases.

Beyond that, there are several previous works \cite{Halpin-Healy1990,HalpinHealy2015} that, by exploiting the connection between the KPZ equation and directed polymers in random media (DPRM), have studied the behavior of this equation on hierarchical lattices. For a certain value of the associated branching parameter, the 1+1 KPZ universality class is recovered, yielding skewness and kurtosis consistent with the TW-GOE distribution. However, these studies also show that there are some differences between hierarchical and Euclidean lattices, suggesting in particular that the $d \to \infty$ limit of KPZ does not correspond to a simple Gumbel distribution as in the limit of infinitely many branches in the DPRM model. In our case, this difference is not relevant either, since the distributions followed by our data are relatively close to the Gaussian and very far from the Gumbel distribution, clearly indicating that Cayley trees are far from representing the infinite-dimensional limit of those equations. This could arise from the fact that the limit distributions obtained in these DPRM studies are not affected by the pathologies we encounter for the Cayley tree, and therefore represent more reliable approaches to the KPZ problem on hierarchical lattices. Actually, if one considers the DPRM model directly on a Cayley tree \cite{Dean2001}, the hierarchical correlations have a considerable effect on the distribution, causing it to deviate from the Gumbel distribution and making it highly nonuniversal.

\section{Acknowledgments}

This work was partially supported by Ministerio de Ciencia, Innovaci\'on y Universidades (Spain), Agencia Estatal de Investigaci\'on (AEI, Spain, 10.13039/501100011033), and European Regional Development Fund (ERDF, A way of making Europe) through Grants No.\ PID2020-112936GB-I00 and No.\ PID2021-123969NB-I00, and by the Junta de Extremadura (Spain) and Fondo Europeo de Desarrollo Regional (FEDER, EU) through Grants No.\ GR21014 and No.\ IB20079. J. M. Marcos is grateful to the Spanish Ministerio de Universidades for a predoctoral fellowship No.\ FPU2021-01334. We have run our simulations in the computing facilities of the Instituto de Computaci\'{o}n Cient\'{\i}fica Avanzada de Extremadura (ICCAEx). The authors thank Alejandro Ortega for discussions about numerical integration techniques for differential equations defined on networks.

\appendix
\section{Local roughness and variance of the average height}\label{appendix_0}

\begin{figure}[t]
\centering
\includegraphics[width=0.495\textwidth]{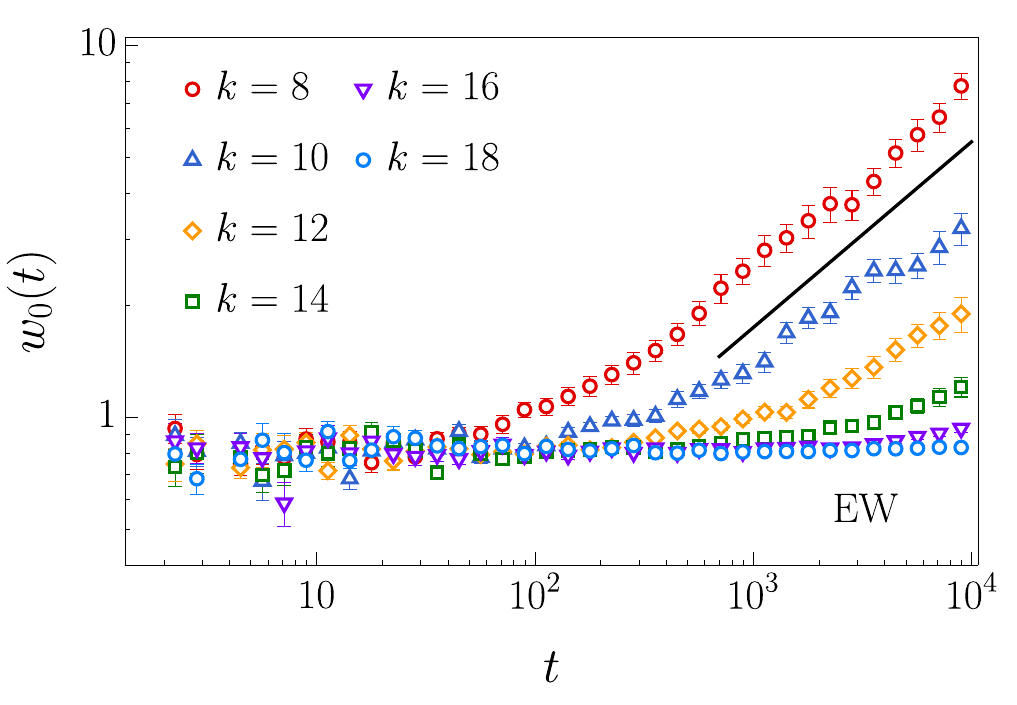}
\includegraphics[width=0.495\textwidth]{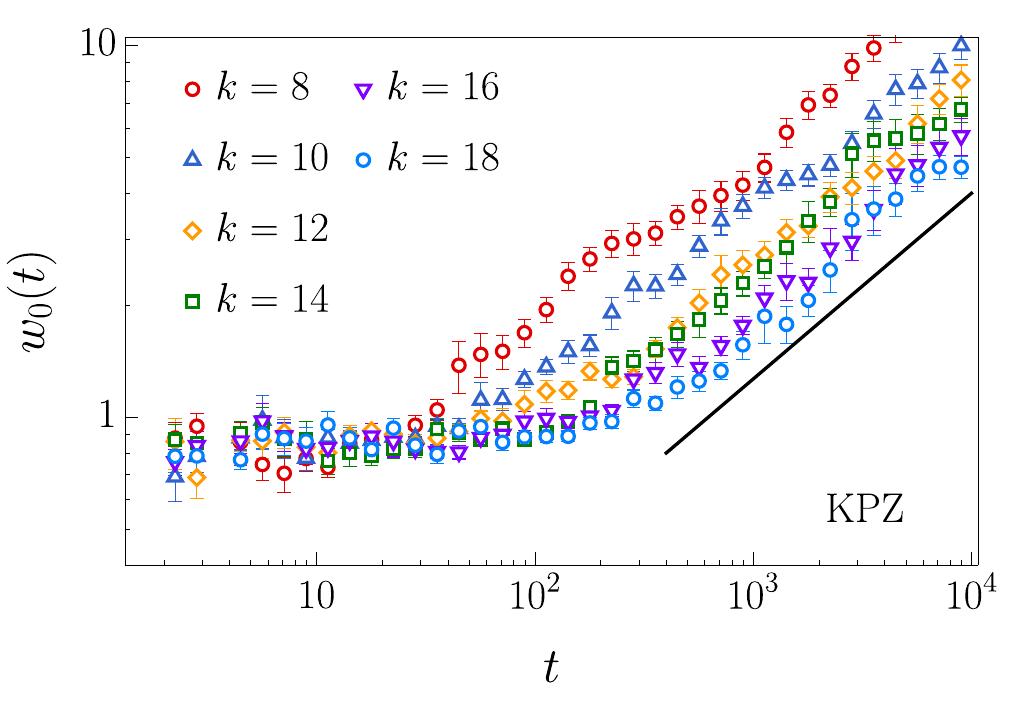}
\caption{Local roughness $w_0(t)$ as a function of time for $q=3$ and several values of $k$ (see legends) for the EW equation (left panel) and the KPZ equation (right panel, where $\lambda=0.5$). As a visual reference, the solid black line in both panels corresponds to $w_0\sim t^{0.5}$. The integration method used was LS.}
\label{fig:w0_integra}
\end{figure}

In this appendix, we examine (local) height fluctuations by now studying the behavior of the \textrm{local} roughness, $w_0(t)$, Eq.\ \eqref{eq:local_roughness}. Figure~\ref{fig:w0_integra} shows the evolution in time of this quantity for various networks with increasing size for both the EW (left panel) and the KPZ equations (right panel). In both cases, the local roughness grows as $w_0\sim t^{1/2}$ after a transient within which its value remains time-independent. This transient becomes longer as the system size increases, especially for the EW case.

\begin{figure}[t]
\centering
\includegraphics[width=0.495\textwidth]{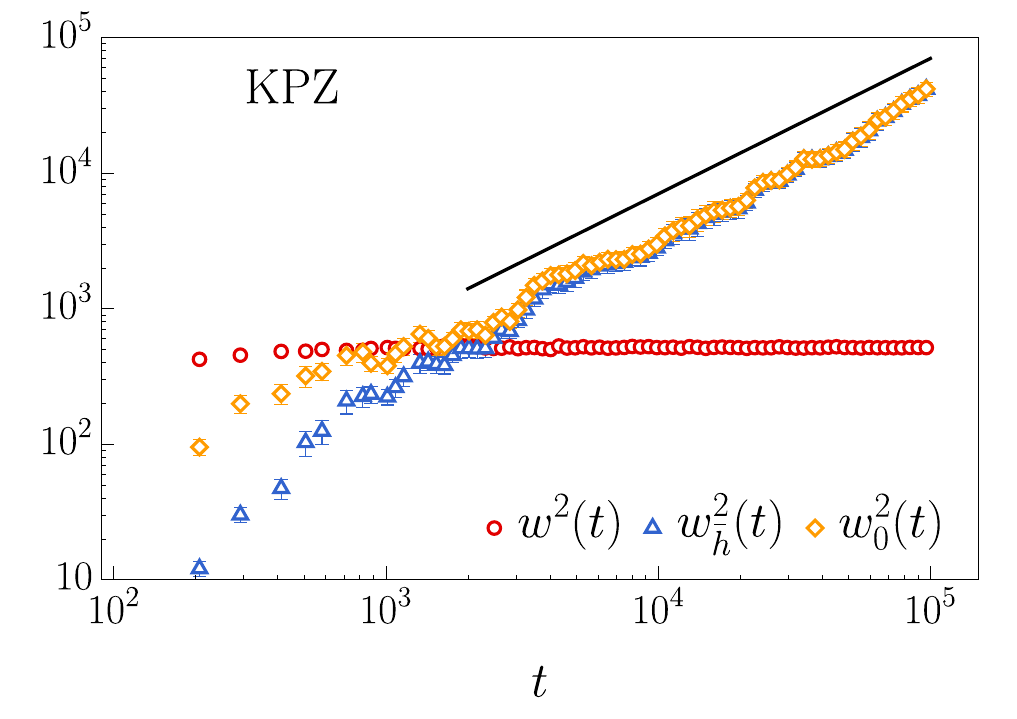}
\includegraphics[width=0.495\textwidth]{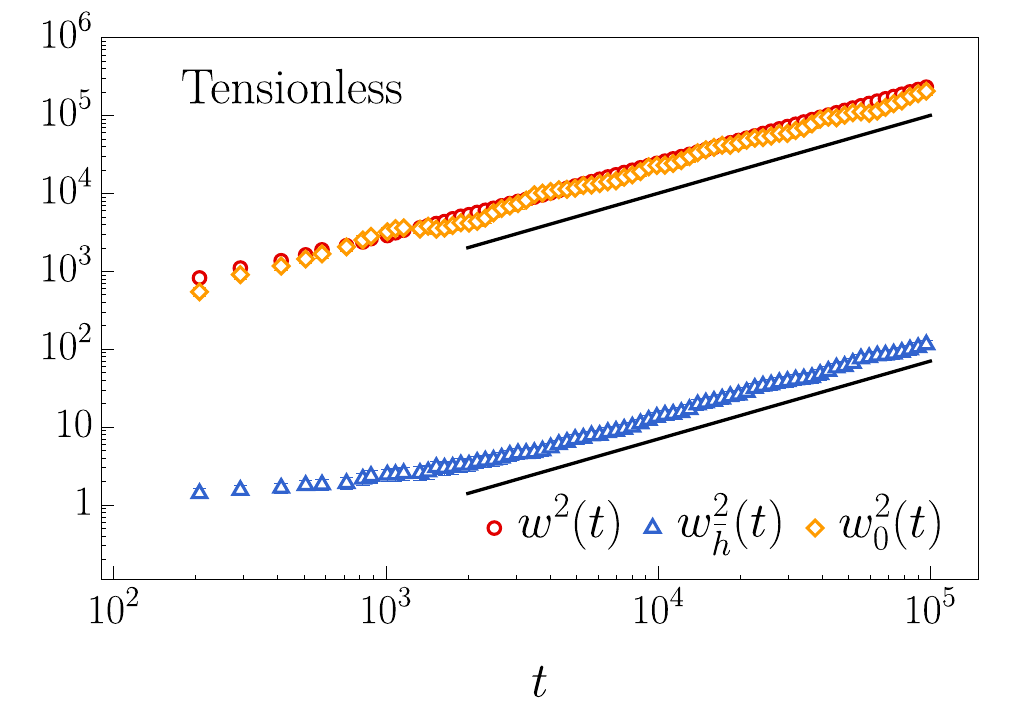}
\caption{Global roughness $w^2(t)$ (red circles), local roughness $w_0(t)$ (yellow diamonds), and the variance of the average height $w_{\bar{h}}(t)$ (blue triangles), as functions of time for one condition of the KPZ equation (left panel) and for one condition of the tensionless KPZ equation (right panel, where $\lambda=3$). As a visual reference, the solid blacks lines corresponds to linear scaling with $t$. In both panels $q=3$ and $k=10$. The integration method used was CI.}
\label{fig:compara_3w}
\end{figure}

In Ref \cite{Oliveira2021}, Oliveira ran simulations for the RDSR model (EW class), two versions of the RSOS model (KPZ class) which differ in how the time is updated, and the BD model (KPZ class), finding flat surfaces $w_0\sim \mathrm{const.}$ for the RDSR model and for one version of the RSOS model (after a short transient), and $w_0\sim t^{1/2}$ for the BD model and the other version of the RSOS models (what Oliveira called the commonest version of the RSOS model, or RSOSc).
Our results are consistent which those reported by Oliveira in \cite{Oliveira2021}, where he explained that saturation was not observed in the BD and the RSOSc models because the variance of the average height was zero due to $\bar{h}$ being deterministic in those models. The argument is that for flat substrates (like a $d$-dimensional regular lattice), thanks to the spatial translation invariance, the one-point height fluctuations can be expressed as
\begin{equation}
    w_0^2=w^2+w_{\overline{h}}^2.
    \label{eq:Oliveira}
\end{equation}
This relation is not exact for a non-flat substrate like the Cayley tree, but a similar trend is expected. As noted by Oliveira, it is well known that $ w_{\overline{h}} \sim t^{1/2} $ in the stationary regime of 1D KPZ and EW systems \cite{Oliveira2021}. In our case, $ \overline{h} $ is never deterministic for the EW, the KPZ, or the tensionless KPZ equations, due to the white noise. As a result, this stochastic contribution is always present when measuring $ w_0 $. We have verified that, for all the equations and conditions studied, the local roughness follows the $ w_0 \sim t^{1/2} $ trend and that the variance of the average height also grows as $ w_{\overline{h}} \sim t^{1/2} $, see Fig.\ \ref{fig:compara_3w}, which displays 
$ w^2(t) $, 
$ w_0^2(t) $, and 
$ w_{\bar{h}}^2(t) $, as directly measured in simulations of the KPZ (left panel) and the tensionless KPZ equations (right panel). The results for the EW equation are very similar to those of the KPZ equation, and are therefore not shown.


Indeed, as seen in Fig.~\ref{fig:compara_3w}, both $ w_0(t) $ and $ w_{\bar{h}}(t) $ grow as $t^{1/2} $, consistently across all simulations performed and regardless of the integration method used. However, for the KPZ equation (as well as for the EW equation), $ w \sim \mathrm{const.} $ because the system conditions chosen in the figure correspond to a saturated global roughness, while $ w_0 \approx w_{\bar{h}} \sim t^{1/2} $. In contrast, for the tensionless KPZ equation, $ w \approx w_0 \sim t^{1/2} \sim w_{\bar{h}} $, but the variance of the average height is significantly smaller as compared with the two other quantities. Hence, for the TKPZ equation the $t^{1/2}$ increase of the global roughness is genuine and not a mere consequence of that of $ w_{\bar{h}}(t) $.

The results for the KPZ and EW equations are consistent with those obtained by Oliveira in Ref.~\cite{Oliveira2021}. The outcome for the tensionless KPZ equation is particularly interesting, especially the $ w^2 \sim t $ growth of global roughness, which does not seem to be followed by saturation to steady state. Such is the behavior expected for the random deposition (RD) model in any dimension \cite{Barabasi1995}. Actually, in \ref{appendix3} we verify that it is indeed obtained for RD on Cayley trees, see Fig.\ \ref{fig:RD1}, which shows that the time evolution of $w^2(t)$, $w_{\bar{h}}^2(t)$, and $w_0^2(t)$ for the RD equation are virtually identical to those of the TKPZ equation, shown in Fig.\ \ref{fig:compara_3w}. 
We do not think that the results of the tensionless case are due to numerical instabilities arising from the integration scheme employed. However, this result may be possibly influenced by boundary effects, such as is the case for the KPZ and EW equations, see \ref{appendix_A}. As Oliveira previously noted (and in our case, from the point of view of the corresponding stochastic equations), the unusual result for EW, for which the upper critical dimension is known to be 2, suggests that the Bethe lattice ---or more precisely, its approximation by Cayley trees--- may not be a suitable substrate for studying the mean-field limit of these universality classes.

\section{Analysis of the growth of layers}\label{appendix_A}

In this appendix, we analyze the growth of each layer within the lattice for each one of the continuum equations under study. Specifically, we calculate the difference between the mean height at the center and at the system boundary, denoted by $\Delta \langle h \rangle$, see Eq.~\eqref{eq:delta_h}. Additionally, we assess the average growth of each layer relative to the global average of the front, denoted $A(s,t)$ and defined in Eq.\ \eqref{eq:capas}. By extending the approach of Ref.\ \cite{Oliveira2021} to our present context, the analysis of these quantities will allow us to understand why the correlation function exhibits jumps in certain cases and why the distribution of fluctuations oscillates in certain occasions.

\begin{figure}[t]
\centering
\includegraphics[width=0.495\textwidth]{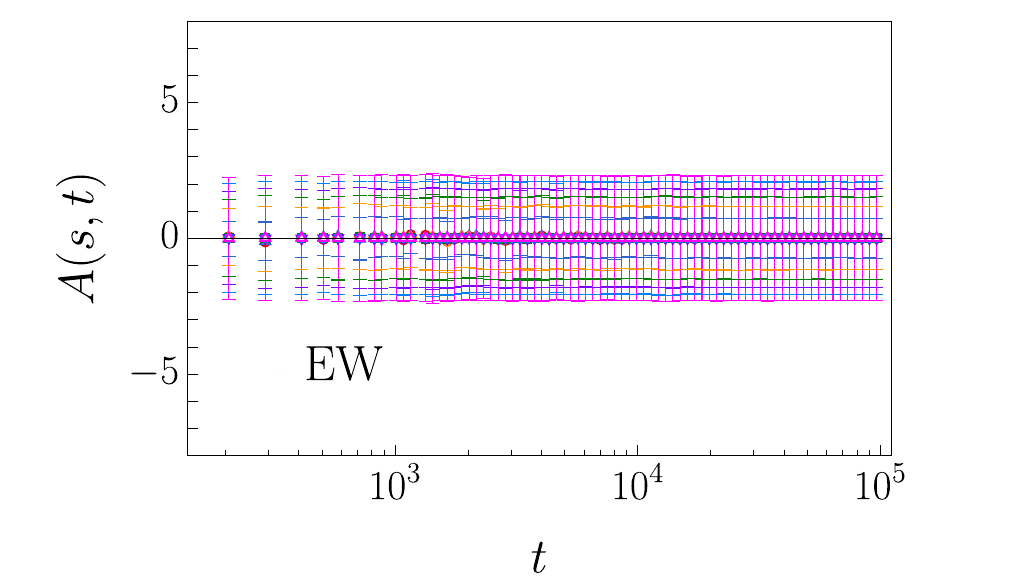}
\includegraphics[width=0.495\textwidth]{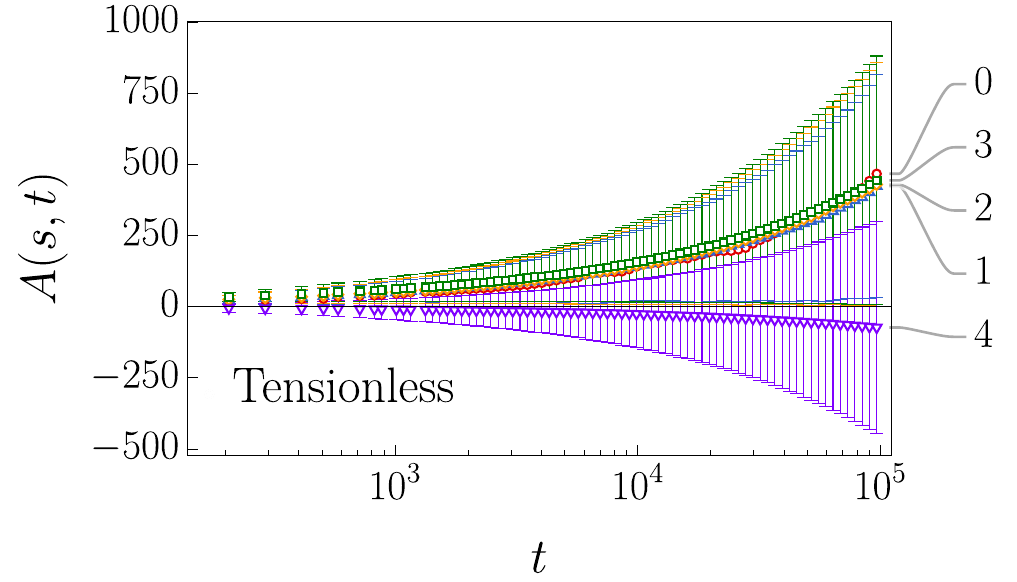}
\caption{Evolution in time of the average of each layer relative to the average front position of the system, $A(s,t)$, for one condition of the EW equation (left panel) and one condition of the tensionless KPZ equation (right panel). In the left panel $q=3$ and $k=8$, while in the right panel $q=6$ and $k=4$. Different colors correspond to different layers. The labels on the right margin of the figure identify each layer value $s$. The integration method used was CI.}
\label{fig:capas}
\end{figure}

\begin{figure}[t]
\centering
\includegraphics[width=0.485\textwidth]{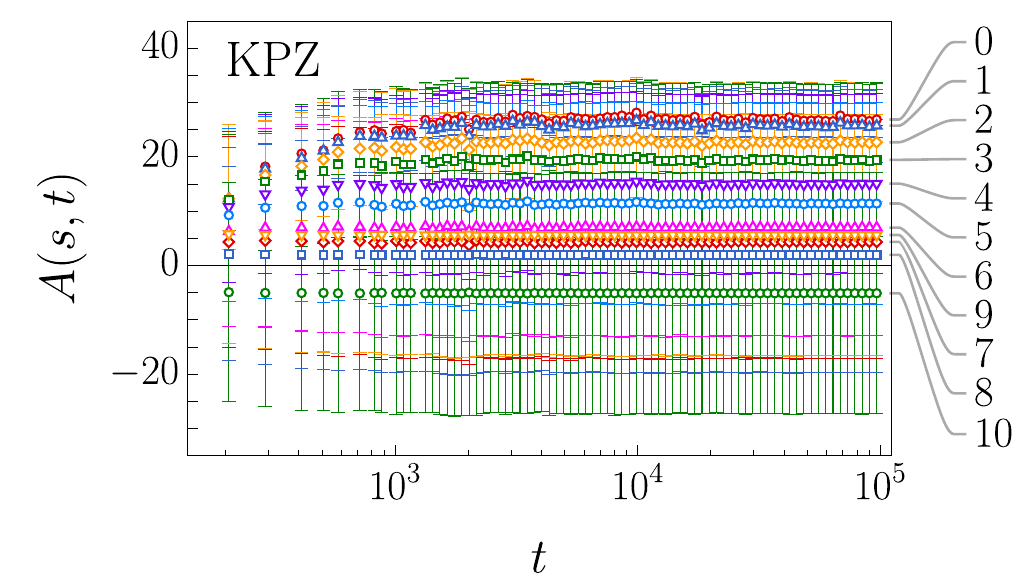}
\includegraphics[width=0.5\textwidth]{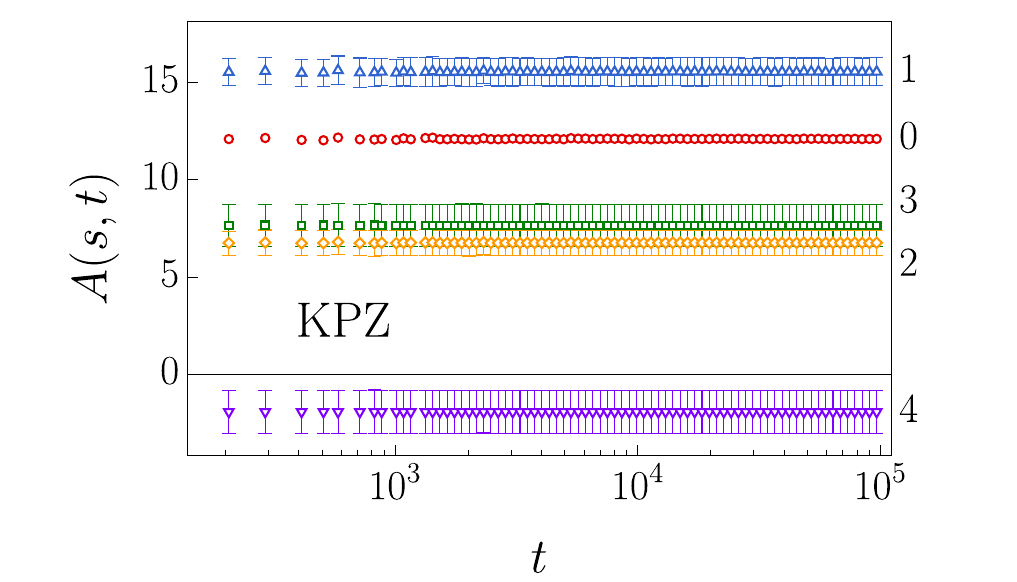}
\caption{Evolution in time of the average of each layer relative to the average front position of the system, $A(s,t)$, for two conditions of the KPZ equation. In the left panel $q=3$ and $k=10$, while in the right panel $q=6$ and $k=4$. The labels on the right margin of the figure identify each layer value $s$ in each case. The integration method used was CI.}
\label{fig:capas2}
\end{figure}

Figure~\ref{fig:capas} shows the evolution in time of $A(s,t)$ for one condition of the EW equation and one condition of the tensionless KPZ equation, while Figure~\ref{fig:capas2} shows analogous results but for two conditions of the KPZ equation. An important preliminary remark is that, in these figures, the error bars  have been estimated differently from the rest of the figures in this work. Instead of using the Jackknife method, we have chosen to define the error as the average (over time and runs) of the standard deviations of each time-box. This choice is helpful in explaining certain features of our system, especially the oscillations observed in the probability distribution of fluctuations in the KPZ case. The presence or absence of overlap between the different layers is crucial for understanding this behavior.

While in the EW equation all layers are grouped around zero, both the KPZ and tensionless KPZ equations exhibit distinctly different behavior for each layer. For the two latter equations, the outermost layer of the system features a mean height which is smaller than the average front height. This behavior remains consistent across all studied conditions. The reason for this is that, since the nodes in the last layer have only one neighbor, their growth driven by the squared-gradient term is significantly smaller than that of the nodes of the inner layers.
This effect propagates layer by layer until reaching the central node of the network, which is typically the one that grows the fastest relative to the average front height. As a result, the closer a layer is to the center of the network, the faster it tends to grow. However, this pattern does not usually hold for the penultimate layer of the system, for which the mean height is often larger than the one immediately closer to the center. We believe this is due to the influence of the outermost layer, which, having a significantly smaller mean than the rest, enhances growth of the penultimate layer. For this reason, when correlations are measured from the center, the penultimate layer in the KPZ case appears to be more correlated than the immediately preceding layers, which are closer to the center.

The same effect occurs also in the simulations of the tensionless KPZ equation. However, in this case, the differences between layers never saturate because there is no relaxation, given the lack of surface tension in the equation. As a result, all inner layers tend to grow very closely together, while the outermost layer increasingly deviates from the rest. This effect leads to the jump observed in the last layer when measuring the correlation function in the case of the tensionless KPZ equation.

By analyzing Fig.~\ref{fig:capas2}, one can understand why oscillations appear in the probability distribution of fluctuations for certain conditions of the KPZ equation, while not in others. In some cases, especially when there are a few layers, as in the right panel of Fig.~\ref{fig:capas2}, there are many gaps (absence of layer overlap) in $A(s,t)$. These gaps represent areas where the fluctuations are highly unlikely. This low density is then reflected when one analyzes the distribution of these fluctuations as, for example, in the right panel of Fig.~\ref{fig:chi}, which corresponds to the same conditions as the right panel of Fig.~\ref{fig:capas2}. In other cases, especially when there are many layers, these tend to overlap with one another. As a result, when analyzing the probability distribution of fluctuations, these fluctuations are observed to decay somewhat continuously (see also the right panel of Fig.~\ref{fig:chi}), following none of our PDF of reference, like Gaussian, or TW.

\begin{figure}[t]
\centering
\includegraphics[width=0.5\textwidth]{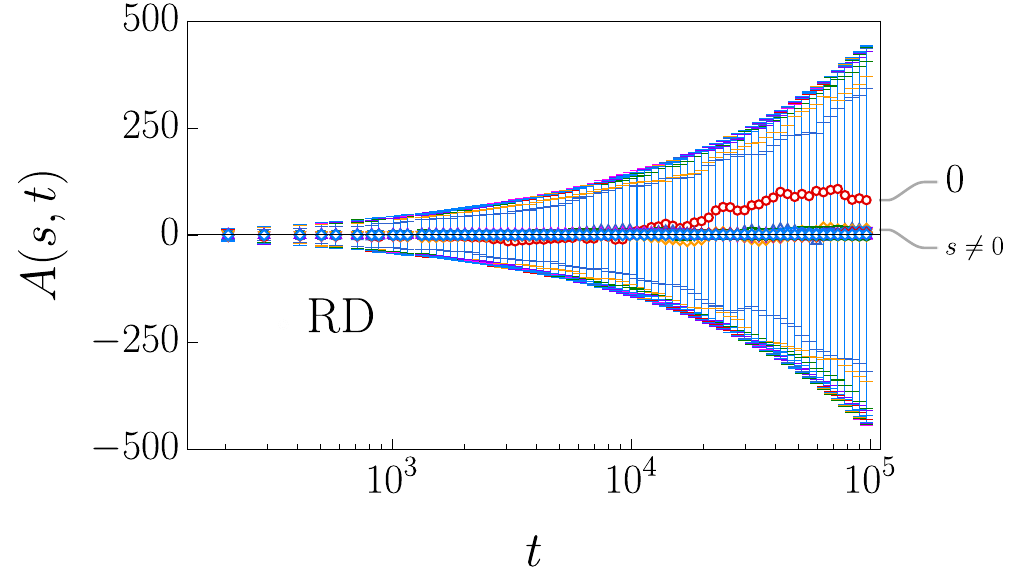}
\caption{Evolution in time of the average of each layer relative to the average front position of the system, $A(s,t)$, for one condition of the RD equation with $q=3$ and $k=12$. The labels on the right margin of the figure identify the central node ($s=0$) and the rest of the layers ($s\ne 0$).}
\label{fig:capas_RD}
\end{figure}

Finally, it is also informative to perform this analysis for the RD equation, whose time evolution for $A(s,t)$ is shown in Fig.~\ref{fig:capas_RD}. As in the EW case, the layer mean height values remain clustered around zero, with the notable exception of the shell at $s = 0$, corresponding to the central node. Since each node in this model evolves independently as a Brownian motion, it is expected that individual sites, such as the central one, may occasionally exhibit significant deviations from the mean. A key distinction is that, due to the absence of relaxation in the RD equation, the error bars increase continuously over time. Recall that these error bars represent the standard deviation of the values within each layer at a given time. This continuous growth is what causes the global roughness to scale as $w(t) \sim t^{1/2}$ in the RD model, in contrast to the EW case, for which the roughness saturates quickly.

The behavior observed in the RD equation closely resembles that of the TKPZ case. In fact, we attribute the growth of the roughness in the TKPZ regime, $w \sim t^{1/2}$, to the fact that the standard deviation of the layer values ---represented by the error bars for $A(s,t)$--- does not saturate over time, similarly to what occurs in the RD model. In this context, we interpret the roughness growth in the TKPZ case as being primarily noise-driven, with the nonlinear term playing a minor role. Nevertheless, a comparison between Figs.~\ref{fig:capas} and ~\ref{fig:capas_RD} shows that the contribution of the nonlinearity is not entirely negligible and does affect the system behavior. In particular, it leads to the emergence of non-zero skewness (i.e., asymmetry) in the height fluctuations which, as mentioned earlier, is a hallmark of the nonlinear term. In contrast, the fluctuations in the RD equation remain Gaussian, see \ref{appendix3}.

\begin{figure}[t]
\centering
\includegraphics[width=0.495\textwidth]{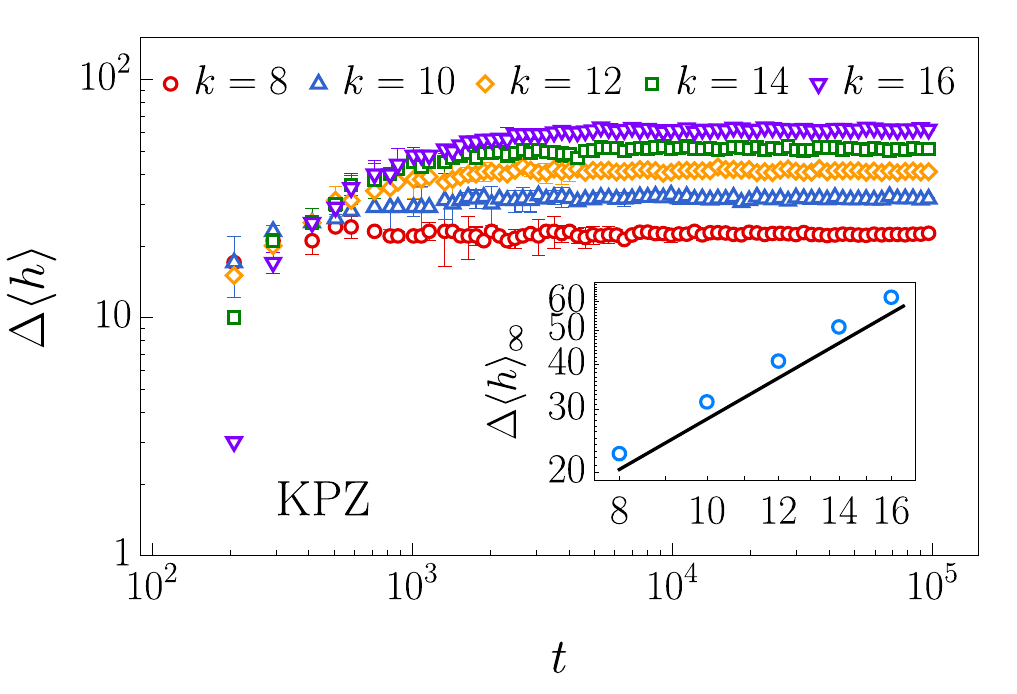}
\includegraphics[width=0.495\textwidth]{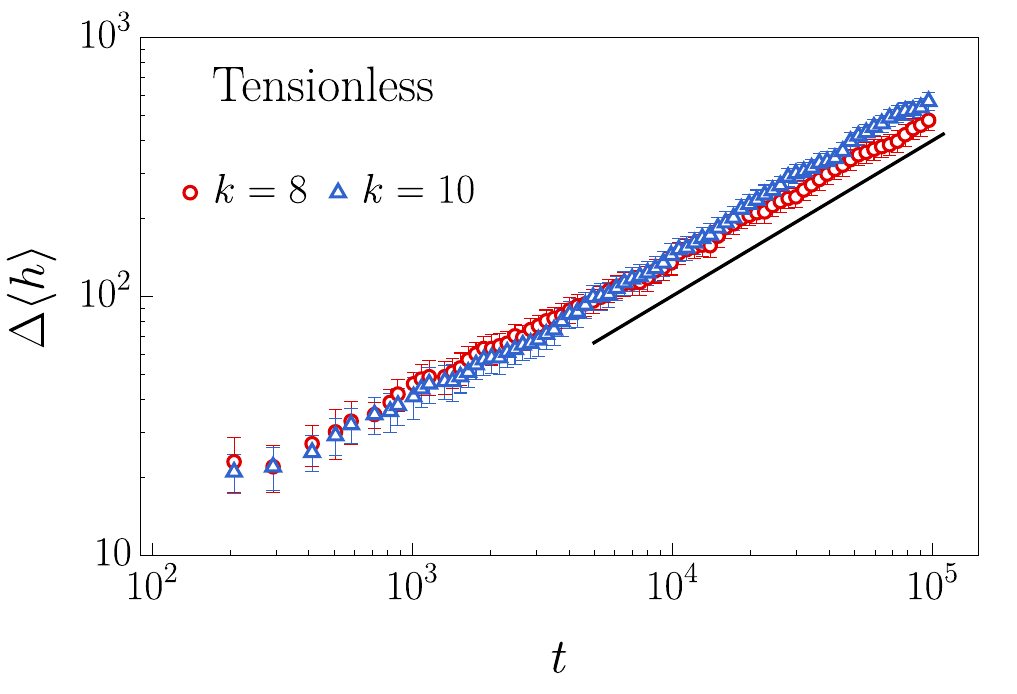}
\caption{Difference between the value of the central node and the average of the last shell $k$, $\Delta\langle h \rangle$ for various system sizes $k$ for the KPZ (left panel) and the tensionless KPZ (right panel) equations. In both panels $q=3$. As a visual reference, the solid black line in the right panel corresponds to $\Delta\langle h \rangle\sim t^{0.6}$. Inset: Saturation value of $\Delta\langle h \rangle_\infty$ versus system size $k$. The solid black line corresponds to $\Delta\langle h \rangle_\infty\sim k^{1.45}$. The integration method used was CI.}
\label{fig:deltaH}
\end{figure}

Following Oliveira \cite{Oliveira2021}, it is also interesting to study the temporal evolution of he difference between the mean height at the center and at the system boundary, $\Delta\langle h\rangle$, defined in Eq.~\eqref{eq:delta_h}. We do not show results for $\Delta\langle h\rangle$ for the EW or RD equations, since this quantity remains essentially zero at all times in both cases. This contrasts with both, the KPZ and the TKPZ equations. Indeed, Fig.~\ref{fig:deltaH} illustrates the nontrivial time evolution of this quantity, as obtained for one condition of the KPZ equation and one condition of the tensionless KPZ equation, for different system sizes. While in the case of the KPZ equation the differences between the central layer and the outermost layer increase over time but eventually saturate, they continue to grow indefinitely in the case of the tensionless KPZ equation, following a power law as $\Delta\langle h \rangle\sim t^{0.6}$. Moreover, in the case of the KPZ equation the saturation value of these differences increases with the system size, following a power law as $\Delta\langle h \rangle_\infty \sim k^{1.45}$. As Oliveira previously pointed out \cite{Oliveira2021}, this is a clear evidence that, in the thermodynamic limit ($k \rightarrow \infty$), the surfaces will be macroscopically curved and, therefore, boundary effects will always prevent an accurate measurement of the global roughness of the system. This is also the case for the tensionless KPZ equation. However, in this case, it is not even necessary to take the thermodynamic limit.

\section{KPZ correlation function vs boundary conditions}\label{appendix2}

\begin{figure}[t]
\centering
\includegraphics[width=0.495\textwidth]{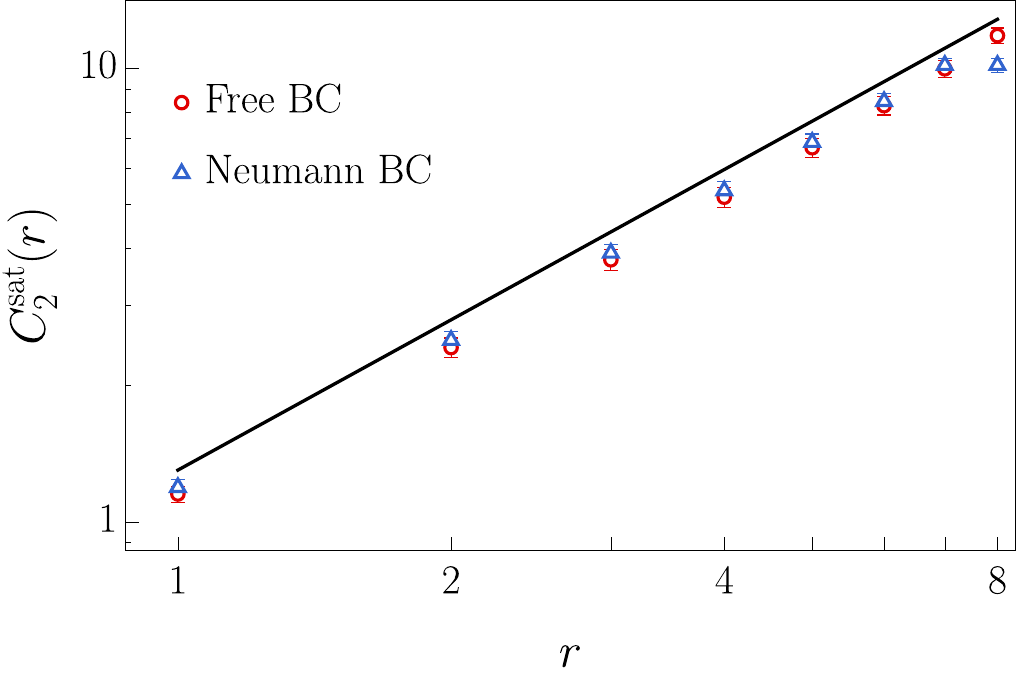}
\includegraphics[width=0.495\textwidth]{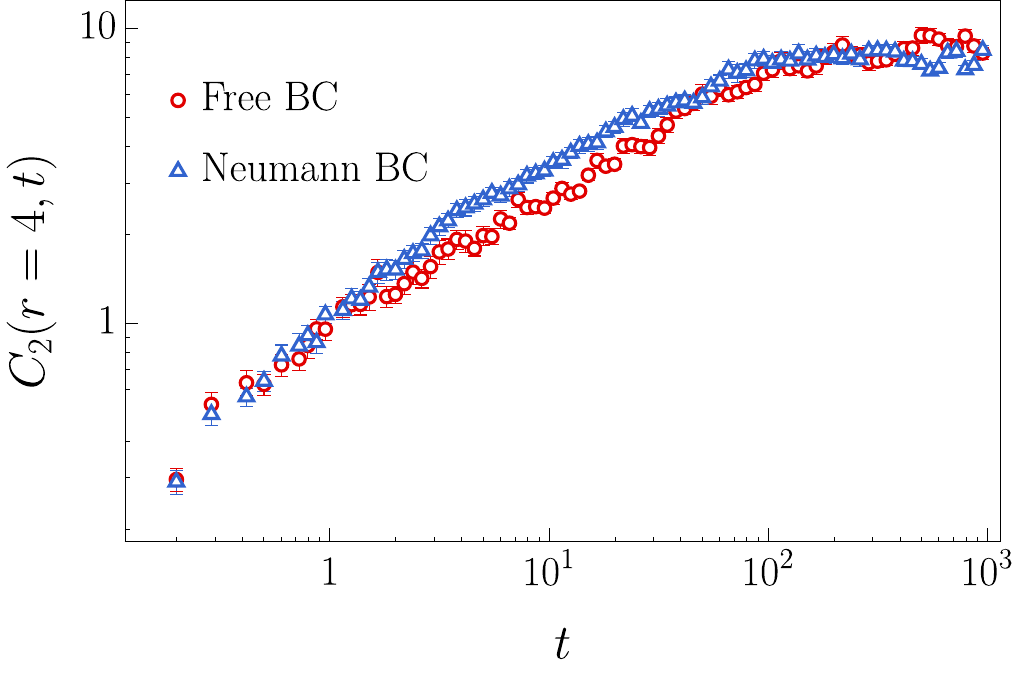}
\caption{Correlation function vs boundary conditions for the KPZ equation. Left panel: Comparison of saturation value of the height difference correlation function, $C^{\rm sat}_2(r)$, 
as a function of $r$ for two different boundary conditions, as in the legend. As a visual reference, the solid black line corresponds to $C_2^{\rm sat}(r)\sim r^{1.1}$. Right panel: Time evolution of the height difference correlation function $C_2(r,t)$ for fixed $r=4$ and the two different boundary conditions. In both panels $q=3$, $k=8$, $\nu=1$, $\lambda=0.5$, and $D=1$. The integration method used was CI.}
\label{fig:comparaC2BC}
\end{figure}

Figure~\ref{fig:comparaC2BC} shows the results of the height-difference correlation function for the KPZ equation for the two BC studied in this work. The left panel displays the saturation value, $C^{\rm sat}(r)$, of $ C_2(r,t) $ as a function of the distance to the center, $ r $, while the right panel illustrates the time evolution of the height-difference correlation function at a fixed distance from the center ($ r=4 $). In the latter panel, it can be observed that the correlation function grows in a very similar way for both cases and reaches the same saturation value. Moreover, at steady state the profile of the correlation function as a function of the distance to the center (left panel) is the same for both boundary conditions, except at the outermost layer. While the Neumann BC forces height values at this layer to equal those of the next-to-last layer, the Free BC allows them to evolve freely, inducing different values of $C^{\rm sat}(r)$ at $r=k$ only.

\section{Results for Random Deposition}\label{appendix3}

\begin{figure}[t]
\centering
\includegraphics[width=0.495\textwidth]{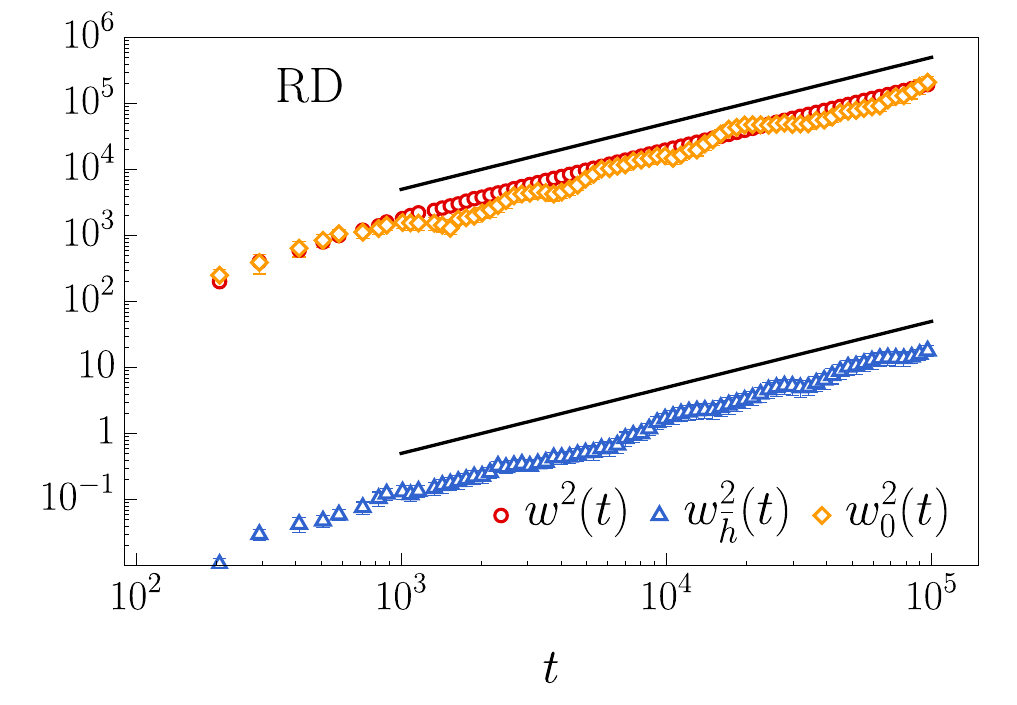}
\includegraphics[width=0.495\textwidth]{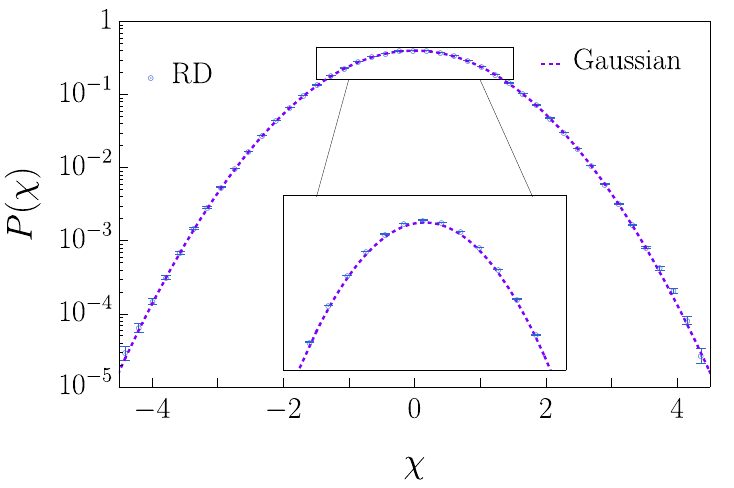}
\caption{Simulations of the RD equation on Cayley trees with $q=3$ and $k=12$, using $D=1$. Left panel: Squared global roughness $w^2(t)$ (red circles), local roughness $w^2_0(t)$ (yellow diamonds), and variance of the average height $w^2_{\bar{h}}(t)$ (blue triangles). As visual references, the solid black lines correspond to linear scaling with $t$. Right panel: Fluctuation histogram of the rescaled height fluctuations, $\chi$, see Eq.~\eqref{eq:chi}. The inset shows a zoom of the boxed area for the central part of the distributions in the $-1.5 < \chi < 1.5$ interval. The dotted purple line correspond to a Gaussian distribution.}
\label{fig:RD1}
\end{figure}

In this appendix, we present the results of simulations of the Random Deposition (RD) equation on Cayley trees. The continuous equation corresponding to this discrete growth model is like Eq.\ \eqref{eq:kpz}, but without the surface tension and nonlinear terms, that is,
\begin{equation}
    \frac{\partial h(\boldsymbol{x},t)}{\partial t}= \eta(\boldsymbol{x},t).
    \label{eq:RD}
\end{equation}
We have performed simulations for different coordination numbers $q$ and numbers of shells $k$ and, as expected, have found no significant differences across these parameters. In all cases, the observed behavior matched the expected behavior for this equation, i.e.\ the results were as found on regular lattices. Specifically, the left panel of Fig.~\ref{fig:RD1} shows the time evolution of $ w^2(t) $, $ w_0^2(t) $, and $ w_{\bar{h}}^2(t) $, while the right panel of Fig.~\ref{fig:RD1} presents histograms of rescaled height fluctuations $\chi$ according to Eq.\ \eqref{eq:chi}. As seen in this figure, the global roughness exhibits the characteristic $w(t) \sim t^{1/2}$ RD growth, while front fluctuations around the mean are Gaussian. 

\begin{figure}[h]
\centering
\includegraphics[width=0.495\textwidth]{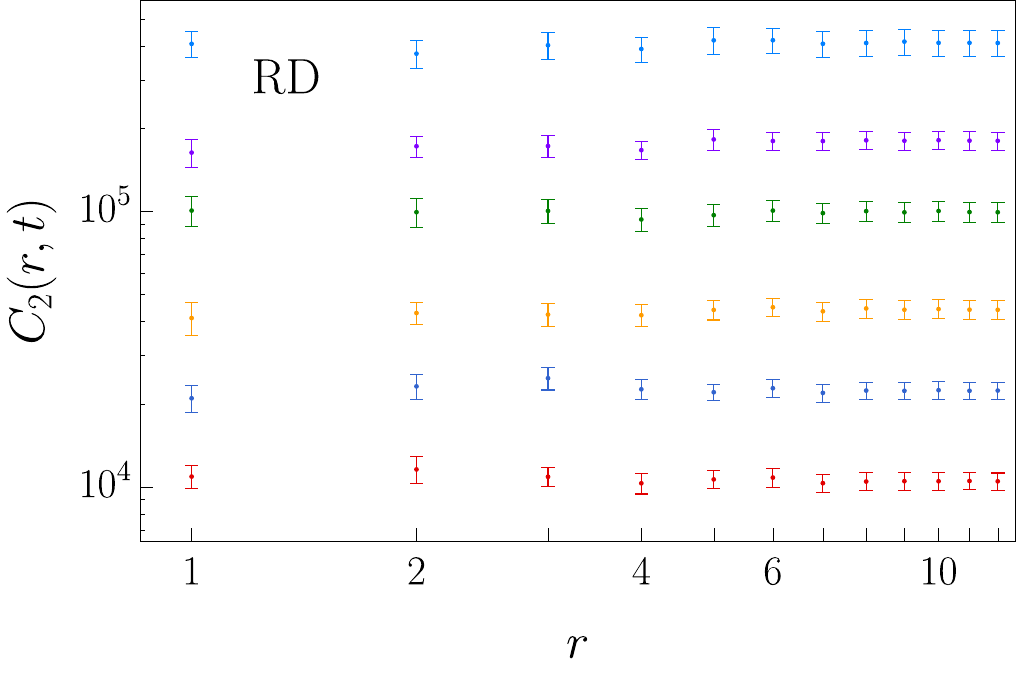}
\includegraphics[width=0.495\textwidth]{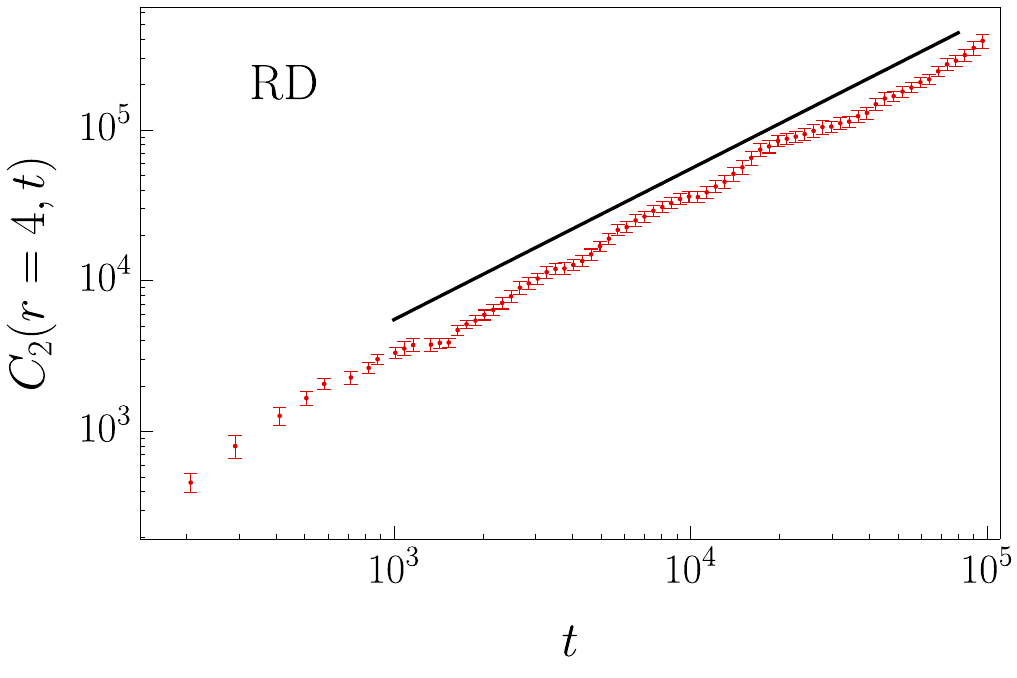}
\caption{Simulations of the RD equation on Cayley trees with $q=3$ and $k=12$, using $D=1$. Left panel: Height-difference correlation function $C_2(r,t)$ as a function of $r$ for $t=25,40,55,70,85$, and $100$. Right panel: Time evolution of the height-difference correlation function $C_2(r,t)$ for fixed $r=4$. As a visual reference, the solid black line corresponds to $C_2(r=4,t)\sim t$.}
\label{fig:RD2}
\end{figure}

Figure~\ref{fig:RD2} shows the results of the height-difference correlation function for the RD equation. The left panel displays $C_2(r,t)$ as a function of the distance to the center, $r$, at various times, where larger values of $C_2(r,t)$ correspond to later times. The right panel illustrates the time evolution of the correlation function at a fixed distance from the center ($r = 4$). As expected, for fixed $t$ the correlation function remains $r$-independent, indicating the absence of space correlations. Moreover, the value of $C_2(r,t)$ increases linearly with time, another hallmark behavior of the RD class.

\section*{References}

\bibliographystyle{iopart-num}
\bibliography{ThinFilm.bib}

\end{document}